\documentclass{jpp}
\usepackage{graphicx,bm,natbib,color,url}

\newcommand{\EQ}{\begin{equation}}
\newcommand{\EN}{\end{equation}}
\newcommand{\EQA}{\begin{eqnarray}}
\newcommand{\ENA}{\end{eqnarray}}
\newcommand{\eq}[1]{(\ref{#1})}
\newcommand{\EEq}[1]{Equation~(\ref{#1})}
\newcommand{\Eq}[1]{equation~(\ref{#1})}
\newcommand{\Eqs}[2]{equations~(\ref{#1}) and~(\ref{#2})}
\newcommand{\eqs}[2]{(\ref{#1}) and~(\ref{#2})}

\newcommand{\Sec}[1]{\S\,\ref{#1}}

\newcommand{\Fig}[1]{figure~\ref{#1}}
\newcommand{\FFig}[1]{Figure~\ref{#1}}

\newcommand{\bra}[1]{\langle #1\rangle}

\newcommand{\meanEMF}{\overline{\mbox{\boldmath ${\cal E}$}} {}}
\newcommand{\meanFF}{\overline{\bm{F}}}
\newcommand{\meanFFFF}{\overline{\mbox{\boldmath ${\cal F}$}} {}}
\newcommand{\meanS}{\overline{S}}
\newcommand{\meanA}{\overline{A}}
\newcommand{\meanB}{\overline{B}}
\newcommand{\meanF}{\overline{F}}
\newcommand{\meanG}{\overline{G}}
\newcommand{\meanJ}{\overline{J}}
\newcommand{\meanK}{\overline{K}}

\newcommand{\meanemf}{\overline{\cal E}}
\newcommand{\meanAA}{\overline{\bm{A}}}
\newcommand{\meanBB}{\overline{\bm{B}}}
\newcommand{\meanJJ}{\overline{\bm{J}}}
\newcommand{\meanKK}{\overline{\bm{K}}}
\newcommand{\meanUU}{\overline{\bm{U}}}
\newcommand{\meanZZ}{\overline{\bm{Z}}}

\newcommand{\meanEE}{\overline{\mbox{\boldmath $E$}}}

\newcommand{\meanrho}{\overline{\rho}}

\newcommand{\eee}{\hat{\mbox{\boldmath $e$}} {}}

\newcommand{\pp}{\hat{\mbox{\boldmath $\phi$}} {}}

\newcommand{\xx}{\mbox{\boldmath $x$} {}}

\newcommand{\zzz}{\mbox{\boldmath $z$} {}}

\newcommand{\uu}{\mbox{\boldmath $u$} {}}
\newcommand{\vv}{\mbox{\boldmath $v$} {}}

\newcommand{\UU}{\mbox{\boldmath $U$} {}}

\newcommand{\bb}{\mbox{\boldmath $b$} {}}

\newcommand{\BB}{\mbox{\boldmath $B$} {}}
\newcommand{\ZZ}{\mbox{\boldmath $Z$} {}}

\newcommand{\AAA}{\mbox{\boldmath $A$} {}}
\newcommand{\aaaa}{\mbox{\boldmath $a$} {}}
\newcommand{\jj}{\mbox{\boldmath $j$} {}}
\newcommand{\JJ}{\mbox{\boldmath $J$} {}}

\newcommand{\kk}{\mbox{\boldmath $k$} {}}

\newcommand{\grav}{\mbox{\boldmath $g$} {}}
\newcommand{\nab}{\mbox{\boldmath $\nabla$} {}}
\newcommand{\OO}{\mbox{\boldmath $\Omega$} {}}
\newcommand{\oo}{\mbox{\boldmath $\omega$} {}}
\newcommand{\ggamma}{\mbox{\boldmath $\gamma$} {}}

\newcommand{\OOmega}{\mbox{\boldmath $\Omega$} {}}

\newcommand{\aalpha}{\mbox{\boldmath $\alpha$} {}}
\newcommand{\eeta}{\mbox{\boldmath $\eta$} {}}
\newcommand{\xxi}{\mbox{\boldmath $\xi$} {}}

\def\urms{u_{\rm rms}}
\def\cP{c_p}
\newcommand{\ii}{{\rm i}}

\newcommand{\dd}{{\rm d} {}}

\def\degr{\hbox{$^\circ$}}
\def\la{\mathrel{\mathchoice {\vcenter{\offinterlineskip\halign{\hfil
$\displaystyle##$\hfil\cr<\cr\sim\cr}}}
{\vcenter{\offinterlineskip\halign{\hfil$\textstyle##$\hfil\cr<\cr\sim\cr}}}
{\vcenter{\offinterlineskip\halign{\hfil$\scriptstyle##$\hfil\cr<\cr\sim\cr}}}
{\vcenter{\offinterlineskip\halign{\hfil$\scriptscriptstyle##$\hfil\cr<\cr\sim\cr}}}}}
\def\ga{\mathrel{\mathchoice {\vcenter{\offinterlineskip\halign{\hfil
$\displaystyle##$\hfil\cr>\cr\sim\cr}}}
{\vcenter{\offinterlineskip\halign{\hfil$\textstyle##$\hfil\cr>\cr\sim\cr}}}
{\vcenter{\offinterlineskip\halign{\hfil$\scriptstyle##$\hfil\cr>\cr\sim\cr}}}
{\vcenter{\offinterlineskip\halign{\hfil$\scriptscriptstyle##$\hfil\cr>\cr\sim\cr}}}}}

\def\Pm{\mbox{\rm Pr}_M}
\def\Rm{R_{\rm m}}

\def\Co{\mbox{\rm Co}}
\def\Beq{B_{\rm eq}}

\def\kf{k_{\rm f}}

\def\nut{\nu_{\rm t}}
\def\etaT{\eta_{\rm T}}
\def\etat{\eta_{\rm t}}
\def\etatz{\eta_{\rm t0}}

\def\EK{E_{\rm K}}
\def\EM{E_{\rm M}}
\def\HM{H_{\rm M}}
\def\EEM{{\cal E}_{\rm M}}

\def\xiM{\xi_{\rm M}}
\def\phiM{\phi_{\rm M}}

\newcommand{\nullvector}{{\bf0}}
\newcommand{\yan}[5]{~ #1~ #5. {\em Astron. Nachr. }{\bf #2}, #3--#4.}
\newcommand{\yanS}[5]{~ #1~ #5 {\em Astron. Nachr. }{\bf #2}, #3--#4.}
\newcommand{\yana}[5]{~ #1~ #5. {\em Astron. Astrophys. }{\bf #2}, #3--#4.}
\newcommand{\yanaN}[4]{~ #1~ #4. {\em Astron. Astrophys. }{\bf #2}, #3.}

\newcommand{\yanaS}[5]{~ #1~ #5 {\em Astron. Astrophys. }{\bf #2}, #3--#4.}

\newcommand{\yajN}[4]{~ #1~ #4. {\em Astron. J. }{\bf #2}, #3.}
\newcommand{\ypnas}[5]{~ #1~ #5. {\em Proc. Natl. Acad. Sci. }{\bf #2}, #3--#4.}

\newcommand{\yass}[5]{~ #1~ #5. {\em Astrophys. Spa. Sci. }
{\bf #2}, #3--#4.}
\newcommand{\ysci}[5]{~ #1~ #5. {\em Science }{\bf #2}, #3--#4.}

\newcommand{\yasr}[5]{~ #1~ #5. {\em Adv. Spa. Res. }{\bf #2}, #3--#4.}
\newcommand{\ysph}[5]{~ #1~ #5. {\em Solar Phys. }{\bf #2}, #3--#4.}

\newcommand{\yjetp}[5]{~ #1~ #5. {\em Sov. Phys. JETP }{\bf #2}, #3--#4.}

\newcommand{\ysov}[5]{~ #1~ #5. {\em Sov. Astron. }{\bf #2}, #3--#4.}

\newcommand{\ymn}[5]{~ #1~ #5. {\em Month. Not. Roy. Astron. Soc. }
{\bf #2}, #3--#4.}

\newcommand{\ynat}[5]{~ #1~ #5. {\em Nature }{\bf #2}, #3--#4.}
\newcommand{\yarXiv}[3]{ ~#1~ #3, arXiv:#2.}
\newcommand{\yjfm}[5]{~ #1~ #5. {\em J. Fluid Mech. }{\bf #2}, #3--#4.}

\newcommand{\ypr}[5]{~ #1~ #5. {\em Phys. Rev. }{\bf #2}, #3--#4.}
\newcommand{\yprl}[5]{~ #1~ #5. {\em Phys. Rev. Lett. }{\bf #2}, #3--#4.}
\newcommand{\yprlN}[4]{~ #1~ #4. {\em Phys. Rev. Lett. }{\bf #2}, #3.}
\newcommand{\yprs}[5]{~ #1~ #5. {\em Proc. Roy. Soc. Lond. }{\bf #2}, #3--#4.}
\newcommand{\yprsa}[5]{~ #1~ #5. {\em Proc. Roy. Soc. Lond. A }{\bf #2}, #3--#4.}
\newcommand{\yptrs}[5]{~ #1~ #5. {\em Phil. Trans. Roy. Soc. }{\bf #2}, #3--#4.}

\newcommand{\yjgr}[5]{~ #1~ #5. {\em J. Geophys. Res. }{\bf #2}, #3--#4.}

\newcommand{\ygrlN}[4]{~ #1~ #4. {\em Geophys. Res. Lett. }{\bf #2}, #3.}
\newcommand{\yrpp}[5]{~ #1~ #5. {\em Rep. Prog. Phys. }{\bf #2}, #3--#4.}
\newcommand{\yjas}[5]{~ #1~ #5. {\em J. Atmosph. Sci. }{\bf #2}, #3--#4.}
\newcommand{\yphlb}[5]{~ #1~ #5. {\em Phys. Lett. B }{\bf #2}, #3--#4.}
\newcommand{\yapj}[5]{~ #1~ #5. {\em Astrophys. J. }{\bf #2}, #3--#4.}
\newcommand{\yapjN}[4]{~ #1~ #4. {\em Astrophys. J. }{\bf #2}, #3.}
\newcommand{\yapjNS}[4]{~ #1~ #4 {\em Astrophys. J. }{\bf #2}, #3.}
\newcommand{\yapjlN}[4]{~ #1~ #4. {\em Astrophys. J. Lett. }{\bf #2}, #3.}
\newcommand{\yapjlNS}[4]{~ #1~ #4 {\em Astrophys. J. Lett. }{\bf #2}, #3.}

\newcommand{\yjppN}[4]{~ #1~ #4. {\em J. Plasma Phys. }{\bf #2}, #3.}
\newcommand{\ypp}[5]{~ #1~ #5. {\em Phys. Plasmas }{\bf #2}, #3--#4.}
\newcommand{\yppN}[4]{~ #1~ #4. {\em Phys. Plasmas }{\bf #2}, #3.}

\newcommand{\ypf}[5]{~ #1~ #5. {\em Phys. Fluids }{\bf #2}, #3--#4.}
\newcommand{\ypfN}[4]{~ #1~ #4. {\em Phys. Fluids }{\bf #2}, #3.}
\newcommand{\yapjs}[5]{~ #1~ #5. {\em Astrophys. J. Suppl. }{\bf #2}, #3--#4.}
\newcommand{\yapjl}[5]{~ #1~ #5. {\em Astrophys. J. Lett. }{\bf #2}, #3--#4.}

\newcommand{\yaraa}[5]{~ #1~ #5. {\em Ann. Rev. Astron. Astrophys. }{\bf #2}, #3--#4.}
\newcommand{\yanf}[5]{~ #1~ #5. {\em Ann. Rev. Fluid Dyn. }{\bf #2}, #3--#4.}
\newcommand{\yoleb}[5]{~ #1~ #5. {\em Orig.\ Life Evol.\ Biosph. }{\bf #2}, #3--#4.}

\newcommand{\yijaS}[5]{~ #1~ #5 {\em Int. J. Astrobiol. }{\bf #2}, #3--#4.}
\newcommand{\ygafd}[5]{~ #1~ #5. {\em Geophys. Astrophys. Fluid Dyn. }{\bf #2}, #3--#4.}
\newcommand{\yphy}[5]{~ #1~ #5. {\em Physica } {\bf #2}, #3--#4.}
\newcommand{\yphyd}[5]{~ #1~ #5. {\em Physica D } {\bf #2}, #3--#4.}
\newcommand{\yjour}[6]{~ #1~ #6. {\em #2} {\bf #3}, #4--#5.}
\newcommand{\yjourS}[6]{~ #1~ #6 {\em #2} {\bf #3}, #4--#5.}
\newcommand{\yjourN}[5]{~ #1~ #5. {\em #2} {\bf #3}, #4.}
\newcommand{\yproc}[7]{~ #1~ #4. In {\em #5} (ed. #6), pp. #2--#3. #7.}

\newcommand{\ybook}[3]{~ #1~ {\em #2}. #3.}
\newcommand{\yprd}[5]{~ #1~ #5. {\em Phys.\ Rev. D }{\bf #2}, #3--#4.}
\newcommand{\yprdN}[4]{~ #1~ #4. {\em Phys.\ Rev. D }{\bf #2}, #3.}
\newcommand{\ypre}[5]{~ #1~ #5. {\em Phys.\ Rev. E }{\bf #2}, #3--#4.}
\newcommand{\ypreN}[4]{~ #1~ #4. {\em Phys.\ Rev. E }{\bf #2}, #3.}
\newcommand{\ypreNS}[4]{~ #1~ #4 {\em Phys.\ Rev. E }{\bf #2}, #3.}

\newcommand{\sana}[3]{ ~#1~ #2. {\em Astron. Astrophys.} (submitted, arXiv:#3).}

\newcommand{\dana}[3]{ ~#1~ #2. {\em Astron. Astrophys.} DOI:#3.}

\newcommand{\sgafd}[2]{ ~#1~ #2. {\em Geophys. Astrophys. Fluid Dyn.} (submitted).}

\newcommand{\sapj}[3]{ ~#1~ #2. {\em Astrophys. J.} (submitted, arXiv:#3).}
\newcommand{\sapjX}[3]{ ~#1~ #2. {\em Astrophys. J.} (submitted, #3).}

\newcommand{\etal}{{\em et al.}}
\def\half{{\textstyle{1\over2}}}

\def\onethird{{\textstyle{1\over3}}}

\newcommand{\G}{\,{\rm G}}

\newcommand{\s}{\,{\rm s}}

\newcommand{\cm}{\,{\rm cm}}

\newcommand{\km}{\,{\rm km}}

\newcommand{\Mm}{\,{\rm Mm}}
\newcommand{\Mx}{\,{\rm Mx}}

\newcommand{\kpc}{\,{\rm kpc}}
\newcommand{\Mpc}{\,{\rm Mpc}}

\newcommand{\Gyr}{\,{\rm Gyr}}

\newcommand{\AU}{\,{\rm AU}}

\newcommand{\const}{\,{\rm const}}

\title[Advances in mean-field theory and its applications]{
Advances in mean-field dynamo theory and applications to astrophysical turbulence}

\author[Axel Brandenburg]{
Axel Brandenburg$^{1,2}$\thanks{
Email address for correspondence: brandenb@nordita.org}}

\affiliation{
$^1$Laboratory for Atmospheric and Space Physics,
JILA, and Department of Astrophysical and Planetary Sciences,
University of Colorado, Boulder, CO 80303, USA\\
$^2$Nordita, KTH Royal Institute of Technology and Stockholm University,
and Department of Astronomy, Stockholm University, SE-10691 Stockholm, Sweden
}

\date{\today,~ $ $Revision: 1.100 $ \!\! $}

\begin{document}

\maketitle

\begin{abstract}
Recent advances in mean-field theory are reviewed and
applications to the Sun, late-type stars, accretion disks, galaxies,
and the early Universe are discussed.
We focus particularly on aspects of spatio-temporal nonlocality,
which provided some of the main new qualitative and quantitative 
insights that emerged from applying the test-field
method to magnetic fields of different length and timescales.
We also review the status of nonlinear quenching and the relation to
magnetic helicity, which is an important observational diagnostic of
modern solar dynamo theory.
Both solar and some stellar dynamos seem to operate in an intermediate
regime that has not yet been possible to model successfully.
This regime is bracketed by antisolar-like differential rotation on one
end and stellar activity cycles belonging to the superactive stars on
the other.
The difficulty in modeling this regime may be related to shortcomings
in modelling solar/stellar convection.
On galactic and extragalactic length scales, the observational constraints
on dynamo theory are still less stringent and more uncertain, but recent
advances both in theory and observations suggest that more conclusive
comparisons may soon be possible also here.
The possibility of inversely cascading magnetic helicity in the early
Universe is particularly exciting in explaining the recently observed
lower limits of magnetic fields on cosmological length scales.
Such magnetic fields may be helical with the same sign of magnetic
helicity throughout the entire Universe.
This would be a manifestation of parity breaking.
\end{abstract}

\section{Introduction}

Hydromagnetic mean-field theory has been instrumental in providing an
early understanding of the oscillatory magnetic field of the Sun with
its 11 year sunspot cycle and the non-oscillatory magnetic field
of the Earth.
This was shown by \cite{SK69a,SK69b} through their numerical
investigations of dynamos in spherical geometry.
These were based on analytical calculations of the $\alpha$ effect and
turbulent magnetic diffusivity a few years earlier \citep{SKR66}.
Now, 50 years later, dynamo theory continues to be an important tool in
many fields of astrophysics and geophysics.
Mean-field theory is also an indispensable tool in predicting the
outcomes of laboratory dynamos
\citep{RRAF02a,RRAF02b,RRAF02c,Forest02,Forest14,Forest15}.
Even now, in the era of large-scale numerical simulations, mean-field
theory provides an important reference to compare against, and to provide
a framework for understanding what happens in the simulations; see,
for example, section~3.4 of \cite{RC14} for attempts to interpret their
simulations using mean-field ideas.
Moreover, numerical simulations have been used to calculate mean-field
transport coefficients such as the $\alpha$ effect and turbulent
magnetic diffusivity without facing the restrictions that analytically
feasible approximations are subjected to.
This has been possible with the development of the test-field method
\citep{Sch05,Sch07}; for a review of this method, see \cite{BCDSHKR10}.
Unfortunately, in spite of significant progress in both numerical and
analytical approaches, there is arguably still no satisfactory model of
the solar dynamo.
The equatorward migration of toroidal magnetic flux belts is not
conclusively understood \citep{SIS06,MT09,Cha10}, and the spoke-like
contours of constant angular velocity, as found through helioseismology
\citep{Schou}, are not well reproduced in simulations.
Simulations have predicted antisolar-like differential rotation
in slowly rotating stars \citep{GYMRW14,KKB14,KKKBOP15} and
nonaxisymmetric global magnetic fields in rapidly rotating stars
\citep{RWBMT90,MBBT95,Viviani18}.
However, the parameters of the transitions from solar-like
to antisolar-like differential rotation and from nonaxisymmetric to
axisymmetric large-scale fields as stars spin down, are not yet well
reproduced in simulations; see Table~5 of \cite{Viviani18}.
The list continues toward larger length scales, from accretion disks
to galactic disks, and even to scales encompassing the entire Universe,
but the observational uncertainties increase in those cases, so the true
extent of agreement between theory and observations is not as obvious
as in the solar and stellar cases.

In this paper, we review the basic deficiencies encountered in modeling
the Sun.
We also highlight some outstanding questions in the applications
of mean-field theory to stars with outer convection zones, to accretion
disks and galaxies, and to the possibility of an inverse cascade of
hydromagnetic turbulence in the early Universe.
We begin by gathering some of the many building blocks of the theory.
Many interesting aspects have emerged over the last 50 years---much of
it became possible through a close interplay between simulations and
analytic approaches.
There is by now a rich repertoire of effects, and it is still not
entirely clear which of them might play a role in the various applications
mentioned above.

\section{Building blocks used in modern mean-field theory}

Mean-field theory can be applied to all the basic equations of
magnetohydrodynamics: the induction equation, the momentum equation,
as well as energy, continuity, and passive scalar equations.
The induction equation is traditionally the best studied one, where
the perhaps most remarkable effects have been discovered.

\subsection{Mean-field induction equation}

In plasmas and other electrically conducting fluids such as liquid metals,
the Faraday displacement current can be omitted compared with the
current density, so the Maxwell equations together with Ohm's law reduce
to the induction equation in the form
\EQ
{\partial\BB\over\partial t}=\nab\times\left(\UU\times\BB-\eta\mu_0\JJ\right)
\label{inductB}
\EN
together with
\EQ
\nab\times\BB=\mu_0\JJ\quad\mbox{and}\quad\nab\cdot\BB=0,
\EN
where $\BB$ is the magnetic field, $\UU$ is the fluid velocity,
$\eta$ is the magnetic diffusivity, $\mu_0$ is the vacuum permeability,
and $\JJ$ is the current density.
At the heart of mean-field theory is a prescription for averaging, denoted
by an overbar.
We then decompose $\UU$ and $\BB$ into mean and fluctuating parts, i.e.,
\EQ
\UU=\meanUU+\uu,\quad
\BB=\meanBB+\bb.
\EN
We choose an averaging procedure which obeys the Reynolds rules, which
state that for any two variables $F=\meanF+f$ and $G=\meanG+g$, we have
\citep{KR80}
\EQ
\overline{\meanF}=\meanF,\quad
\overline{f}=0,\quad
\overline{F+G}=\meanF+\meanG,\quad
\overline{\meanF\,\meanG}=\meanF\,\meanG,\quad
\overline{\meanG f}=\meanG.
\EN
These rules imply that
\EQ
\overline{\UU\times\BB}=\meanUU\times\meanBB+\overline{\uu\times\bb}
\EN
and
\EQ
(\UU\times\BB)'=\meanUU\times\bb+\uu\times\meanBB
+\uu\times\bb-\overline{\uu\times\bb},
\label{UxBprime}
\EN
where the prime denotes the fluctuating part.\footnote{
In the following, we continue using the lowercase symbols $\uu$ and $\bb$
instead of $\UU'$ and $\BB'$ to denote fluctuations of $\UU$ and $\BB$.}
The mean-field induction equation is thus given by
\EQ
{\partial\meanBB\over\partial t}=\nab\times\left(
\meanUU\times\meanBB+\overline{\uu\times\bb}-\eta\mu_0\meanJJ\right)
\label{dmeanBBdt}
\EN
together with $\nab\times\meanBB=\mu_0\meanJJ$ and $\nab\cdot\meanBB=0$.

The next important step here is the calculation of the mean electromotive
force $\meanEMF=\overline{\uu\times\bb}$.
One often makes the assumption of an instantaneous and local response
in terms of $\meanBB$ of the from \citep{KR80}
\EQ
\meanemf_i=\meanemf_i^{(0)}+\alpha_{ij}\meanB_j+\eta_{ijk}\meanB_{j,k}
\quad\quad\mbox{(local \& instantaneous)},
\label{LocalInstantaneous}
\EN
where the comma in $\meanB_{j,k}$ denotes partial differentiation and
$\meanEMF^{(0)}$ is a nonvanishing contribution to the mean electromotive
force for $\meanBB=\nullvector$; see \cite{BR13} for examples of terms
proportional to the local angular velocity and the cross helicity
$\overline{\uu\cdot\bb}$.
This is also known as the Yoshizawa effect \citep{YY93,Yok13}.
Since the Yoshizawa effect leads to a growth even without a formal
large-scale seed magnetic field, it is sometimes referred to as a
turbulent battery effect \citep{BU98}.
It is generally caused by the presence of cross helicity, which
can be generated when a mean magnetic field is aligned with the direction
of gravity \citep{RKB11}.
Originally, \cite{YY93} discussed applications primarily to accretion
and galactic disks, but in recent year, applications to solar and stellar
dynamos have also been discussed \citep{Pip11,YSPH16}.

Let us now return to the other two terms in \Eq{LocalInstantaneous}.
To find expressions for $\alpha_{ij}$ and $\epsilon_{ijk}$, one
has to compute $\meanEMF=\overline{\uu\times\bb}$.
We postpone the discussion of the evolution of $\uu$ until \Sec{TauApprox}
and consider here only the evolution equation for $\bb$, which is obtained
by subtracting \Eq{dmeanBBdt} from \Eq{inductB} and using \Eq{UxBprime}.
This yields
\EQ
{\partial\bb\over\partial t}=\nab\times\left(\meanUU\times\bb+\uu\times\meanBB
+\uu\times\bb-\overline{\uu\times\bb}-\eta\mu_0\jj\right).
\label{inductb}
\EN
The term $\uu\times\bb-\overline{\uu\times\bb}$ is nonlinear in the
fluctuations.
It is important in all cases of practical interest, such as turbulent and
steady flows at large magnetic Reynolds numbers (low magnetic diffusivity)
and will be discussed further in \Sec{TauApprox}.
In the second-order correlation approximation (SOCA), however, one
neglects this term.
This is permissible not only when $\eta$ is large (small magnetic Reynolds
number), but also when the correlation time is short.
In these cases, the nonlinear term is overpowered either by the diffusion
term, $\nab\times(-\eta\mu_0\jj)=\eta\nabla^2\bb$ (for $\eta=\const$)
on the right-hand side, or by the $\partial\bb/\partial t$ term on the
left-hand side of \Eq{inductb}.
Neglecting now also the effects of a mean flow ($\meanUU=\nullvector$)
and assuming incompressibility ($\nab\cdot\uu=0$), SOCA yields 
\EQ
\left({\partial\over\partial t}-\eta\nabla^2\right)\bb
=\meanBB\cdot\nab\uu-\uu\cdot\nab\meanBB.
\label{inductb_approx}
\EN
This equation can be solved using the Green's function for the heat
equation which, in Fourier space with frequency $\omega$ and wavenumber
$k$, is given by $(-\ii\omega+\eta k^2)^{-1}$.
When applied to calculating $\meanEMF$, this corresponds
in the end to a multiplication by a correlation time $\tau$
\citep[see details in][]{Mof70,Mof78,KR80}.
Thus, we have
\EQ
\meanemf_i=\tau\epsilon_{ijk}\left(\overline{u_j u_{k,l}}\,\meanB_l 
-\overline{u_j u_l}\,\meanB_{k,l}\right)
\equiv\alpha_{il}\meanB_l+\eta_{ikl}\meanB_{k,l},
\label{demfidt}
\EN
where $\alpha_{il}=\tau\epsilon_{ijk}\overline{u_j u_{k,l}}$ is
the $\alpha$ tensor and $\eta_{ikl}=\tau\overline{u_j u_l}$ has
a part that contributes to turbulent magnetic diffusion.

To give an explicit example, let us first discuss the isotropic idealization.
In that case, $\alpha_{il}$ and $\eta_{ikl}$ must be isotropic tensors.
The only isotropic tensors of ranks two and three are $\delta_{il}$ and
$\epsilon_{ikl}$, respectively.
Thus, we write $\alpha_{ij}=\alpha\delta_{ij}$ and
$\eta_{ijk}=\etat\epsilon_{ijk}$, where $\alpha$ is a pseudoscalar and
$\etat$ is a regular scalar (the turbulent magnetic diffusivity).
For sufficiently large magnetic Reynolds numbers (low magnetic
diffusivity) the two are given approximately by what we call their
reference values $\alpha_0$ and $\etatz$, defined through
\EQ
\alpha_0\equiv-\onethird\tau\overline{\oo\cdot\uu},\quad\quad
\etatz\equiv\onethird\tau\overline{\uu^2},
\label{alpha0Formula}
\EN
where $\tau\approx(\urms\kf)^{-1}$ is the turbulent turnover time,
$\urms=(\overline{\uu^2})^{1/2}$ is the rms velocity of the fluctuations,
$\kf$ is the wavenumber of the energy-carrying eddies,
and $\oo=\nab\times\uu$ is the fluctuating vorticity.
Since $\epsilon_{ijk}\partial_k\meanB_j=-(\nab\times\meanBB)_i$,
i.e., with a minus sign, the mean electromotive force is given by
$\meanEMF=\meanEMF^{(0)}+\alpha\meanBB-\etat\mu_0\meanJJ$.
The approximations used to obtain $\alpha\approx\alpha_0$ and
$\etat\approx\etatz$ only hold for magnetic Reynolds
numbers, $\Rm=\urms/\eta\kf$, that are larger than unity.
For smaller values of $\Rm$, $\alpha$ and $\etat$ increase linearly
with $\Rm$.
It must also be emphasized that \cite{SBS08} found \Eq{alpha0Formula}
to be valid for turbulent flows where $\tau$ is not small.

In practice, astrophysical turbulence is always driven by some
kind of instability.
Highly supercritical Rayleigh-Ben\'ard convection is an example where
the turbulence is inhomogeneous and therefore also anisotropic.
The Bell instability \citep{Bell04} is driven by a cosmic-ray
current, producing anisotropic turbulence.
In these cases, anisotropy and inhomogeneity of the turbulence are
characterized by one preferred direction, $\eee$.
This can be used to simplify the complexity of the expression for
$\meanEMF$ to
\EQA
\meanEMF_\perp&=&\alpha_\perp\meanBB_\perp-\eta_\perp\mu_0\meanJJ_\perp-\kappa_\perp\meanKK_\perp
              +\gamma\eee\times\meanBB_\perp-\delta\eee\times\mu_0\meanJJ_\perp-\mu\eee\times\meanKK_\perp,\\
\meanEMF_\|&=&\alpha_\|\meanBB_\|-\eta_\|\mu_0\meanJJ_\|-\kappa_\|\meanKK_\|,
\ENA
with only nine coefficients instead of $9+27=36$ for the
full rank two and three tensors.
Here, $\meanK_i=\half(\meanB_{i,j}+\meanB_{j,i})\hat{e}_j$ is a
vector characterizing the symmetric part of $\meanB_{i,j}$, while
$\meanJ_i=-\half\epsilon_{ijk}\meanB_{j,k}$ characterizes its
antisymmetric part.
\cite{BRK12} have determined all these coefficients for
their forced turbulence simulations using rotation, stratification,
or both as preferred directions of their otherwise isotropically
forced turbulence.

Instead of repeating what has been discussed and reviewed extensively
in the literature \citep{Mof78,Par79,KR80,ZRS83,Rae90,RoS92,BS05},
we first focus on aspects that may turn out to be rather important,
namely nonlocality in space and time.
Both are long known to exist \citep{Rae76}, but only recently has their
importance become apparent.
This may be important in solving some of the long-standing problems in
astrophysical magnetism.
Next, we discuss the status of $\alpha$ quenching and the relation to
magnetic helicity fluxes, which is an important diagnostics in solar
physics \citep{KMRS02,KMRS03}.

\subsection{Nonlocality: when scale separation becomes poor}
\label{NonlocalityScaleSeparation}

One often makes the assumption of a separation of scales between the
scale of large-scale magnetic fields and the scale of the energy-carrying
eddies or fields, which are referred to as small-scale fields.
In real applications, this is often not well justified.
Think, for example, of the convective downflows extending over a
major part of the convection zone, or of the possibility of giant cell
convection \citep{Miesch08}.
When scale separation does indeed become poor, one cannot adopt the
local and instantaneous connection used in \Eq{LocalInstantaneous}, but
one has to resort to the integral kernel formulation,
\EQ
\meanemf_i(\xx,t)=\meanemf_i^{(0)}+\int\!\!\!\int{\cal K}_{ij}(\xx,\xx',t,t')\,
\meanB_j(\xx',t')\,\dd^3\xx'\,\dd t',
\label{Kernel}
\EN
as was explained by \cite{Rae76},
It is convenient to retain a formulation similar to that of
\Eq{LocalInstantaneous}, and write
\EQ
\meanemf_i=\meanemf_i^{(0)}+\hat\alpha_{ij}\circ\meanB_j
+\hat\eta_{ijk}\circ\meanB_{j,k}
\quad\quad\mbox{(nonlocal with memory)},
\label{Memory}
\EN
where the symbol $\circ$ denotes a convolution and $\hat\alpha_{ij}$
and $\hat\eta_{ijk}$ are integral kernels.
This all sounds troublesome, because a convolution over time requires
keeping the full history of $\meanB_j(\xx',t')$ over all past times $t'$
at all positions $\xx'$.
However, there is actually a simple {\em approximation} which captures
the {\em essential} effects of nonlocality in space and time.
This will be explained below.

As will become clear in the next section. the importance of spatial
nonlocality lies in the fact that it prevents the unphysical occurrence
of small-scale structures in a mean-field dynamo.
Nonlocality in time is also important, because it can lead to {\em new} dynamo
effects of their own, as will also be explained in a moment.

Let us now discuss the term $\meanEMF^{(0)}$, whose relation to
nonlocality has not previously been emphasized.
\cite{BR13} discussed contributions to $\meanemf_i^{(0)}$ of the
form $c_\Omega\Omega_i$, where $\OO$ is the angular velocity and
$c_\Omega$ is a dynamo coefficient proportional to the cross helicity,
$\overline{\uu\cdot\bb}$.
A similar contribution is of the form $c_\omega\omega_i$, where
$\oo$ is the local vorticity.
If written in this form, it becomes plausible that these terms generalize
to $c_\Omega\circ\Omega_i$ or $c_\omega\circ\omega_i$, and that it is
thus no exception to the treatment as a convolution.

\subsection{A practical tool for capturing the essence of nonlocality}
\label{TheTrick}

A decisive step in arriving at an approximate expression for the
nonlocality in space and time was the development of the test-field method
for calculating turbulent transport coefficients \citep{Sch05,Sch07}.
This is a method for calculating $\alpha$ effect, turbulent diffusivity,
and other turbulent transport coefficients for arbitrary mean magnetic
fields.
It turned out that test fields of high spatial wavenumber $k$ tend to
result in transport coefficients that are decreased approximately like
a Lorentzian proportional to $1/(1+k^2/\kf^2)$; see \cite{BRS08}.
Likewise, it was found that rapid variations in time proportional to
$e^{-\ii\omega t}$ with frequency $\omega$ lead to a reduced and modified
response along with a frequency-dependent delay; see \cite{HB09}.
In frequency space, the corresponding response kernel was found to be
of the form $1/(1-\ii\omega\tau)$, where $\tau$ is a typical response
or correlation time, namely the $\tau\approx(\urms\kf)^{-1}$ stated above.
Thus, no new unknown physical parameters enter and everything is in
principle known.

We recall that a convolution in space and time, as expressed by
\Eqs{Kernel}{Memory}, corresponds to a multiplication in wavenumber and
frequency space.
Furthermore, the combined $k$ and $\omega$ dependence of our kernels
was found to be proportional to $1/(1-\ii\omega\tau+k^2/\kf^2)$.
This was verified empirically with the test-field method \citep{RB12}.
Thus, we have
\EQ
\left(1-\ii\omega\tau+k^2/\kf^2\right)\meanemf_i=\meanemf_i^{(0)}
+\tilde\alpha_{ij}\meanB_j+\tilde\eta_{ijk}\meanB_{j,k}.
\label{EMFomk}
\EN
This can easily be expressed in real space as an evolution equation
for $\meanEMF$ along with a diffusion term,
\EQ
{\partial\meanemf_i\over\partial t}={1\over\tau}\left(\meanemf_i^{(0)}
+\alpha_{ij}^{(0)}\meanB_j+\eta_{ijk}^{(0)}\meanB_{j,k}-\meanemf_i\right)
+\kappa_{\cal E}\nabla^2\meanemf_i,
\label{dEMFdt}
\EN
where $\kappa_{\cal E}=(\tau\kf^2)^{-1}$ is an effective diffusivity for $\meanEMF$,
and $\alpha_{ij}^{(0)}$ and $\eta_{ijk}^{(0)}$ are now no longer integral
kernels, but just functions of space and time (in addition of course to other
parameters of the system itself).
So, instead of a cumbersome convolution, we now have instead
a much simpler differential equation in space and time.
In other words, instead of an instantaneous and local response, as in
\Eq{LocalInstantaneous}, we now have an evolution equation along with
a stabilizing turbulent diffusion term, which is computationally very
convenient.
Note that now the $\meanemf_i^{(0)}$ term is automatically treated as
a convolution, too.
This is, as argued above, to be expected and could be important
provided the vorticity vector, which would enter this term,
is space- and time-dependent.

\subsection{Tau approach and physical reality of an
evolution equation for $\meanEMF$}
\label{TauApprox}

The physical reality of an evolution equation for $\meanEMF$ was first
proposed by \cite{BF02} as a natural consequence of retaining the time
derivative introduced in the $\tau$ approximation---or better
$\tau$ ``approach'', because it is not a controlled approximation.
To understand the connection with an evolution equation for $\meanEMF$,
let us briefly review the essence of this approach.
Unlike SOCA, where one needs only the evolution \Eq{inductb} for $\bb$,
we now also need an evolution equation for $\uu$.
Here we assume it to be mainly governed by the Lorentz force, $\JJ\times\BB$,
\EQ
{\partial\uu\over\partial t}=\meanJJ\times\bb+\jj\times\meanBB+
\jj\times\bb-\overline{\jj\times\bb}+\nu\nabla^2\uu+...,
\label{dudt_approx}
\EN
where the ellipsis indicates additional terms such as the pressure
gradient and the advection term that are here omitted.
Next, we calculate
\EQ
{\partial\meanEMF\over\partial t}
=\overline{\uu\times\dot{\bb}}+\overline{\dot{\uu}\times\bb},
\label{TwoTerms}
\EN
where the dots on $\uu$ and $\bb$ indicate partial derivatives with
respect to time.
Retaining only the term resulting from tangling of $\meanBB$, we have
\EQ
{\partial\meanemf_i\over\partial t}
=\epsilon_{ijk}\left(\,\overline{u_j\meanB_l u_{k,l}+\meanB_l b_{j,l}b_k}\,\right)+...
=\left(\alpha_{il}^{\prime\rm K}+\alpha_{il}^{\prime\rm M}\right)\meanB_l+...\;,
\label{demfidt}
\EN
where $\alpha_{il}^{\prime\rm K}=\epsilon_{ijk}\overline{u_j u_{k,l}}$
and   $\alpha_{il}^{\prime\rm M}=\epsilon_{ijk}\overline{b_{j,l} b_k}$
are proportional to the kinetic and magnetic $\alpha$ effects
(the actual $\alpha$ effects will be without primes)
and commas denote partial differentiation.
Their traces are
$\alpha_{ii}^{\prime\rm K}=\epsilon_{ijk}\overline{u_j u_{k,i}}=-\overline{\oo\cdot\uu}$ and
$\alpha_{ii}^{\prime\rm M}=\epsilon_{ijk}\overline{b_{j,i} b_k}=\overline{\jj\cdot\bb}$,
but the essential part for our discussion lies in the ellipsis.
In the $\tau$ approach, one assumes that triple correlations
resulting from the nonlinearities can be approximated by the
quadratic correlation as $-\meanEMF/\tau$ on the right-hand side
of \Eq{demfidt}, where $\tau$ is a relaxation (or correlation)
time, which lent its name to this approach.
This leads directly to $(1+\tau\partial_t)\meanEMF=\alpha\meanBB+...$,
where $\alpha=\onethird\tau(\alpha_{ii}^{\prime\rm
K}+\alpha_{ii}^{\prime\rm M})$ in the isotropic case and the ellipsis
denotes higher order derivatives giving rise to turbulent diffusion,
etc, which are still being captured both by SOCA and the $\tau$ approach,
but that were omitted for the sake of a simpler presentation.

\cite{BF03} applied the idea of retaining the time derivative introduced
in the $\tau$ approach to the case of passive scalar transport,
where the instantaneous Fickian diffusion approximation is replaced by
a telegrapher's equation.
The physical reality of the telegrapher's equation in turbulent transport
was subsequently confirmed using numerical simulations \citep{BKM04}.
It turns the parabolic diffusion equation into a damped wave equation
with a wave speed that is the turbulent rms velocity in the direction
of the wave.
For large turbulent diffusivities, this approach also avoids uncomfortably
short timesteps in numerical solutions.
Examples where this approach was used include cosmic ray transport in the
interstellar medium \citep{SBMS06} and field-aligned thermal conduction
in the solar corona \citep{Rem17}.
A particular effect of interest is that of a spiral forcing of the
dynamo coefficients, which was found to result also in a shift of the
mean-field spiral response by a factor of the order of $\Omega\tau$;
see \cite{CSS13}.

\begin{figure}\begin{center}
\includegraphics[width=\textwidth]{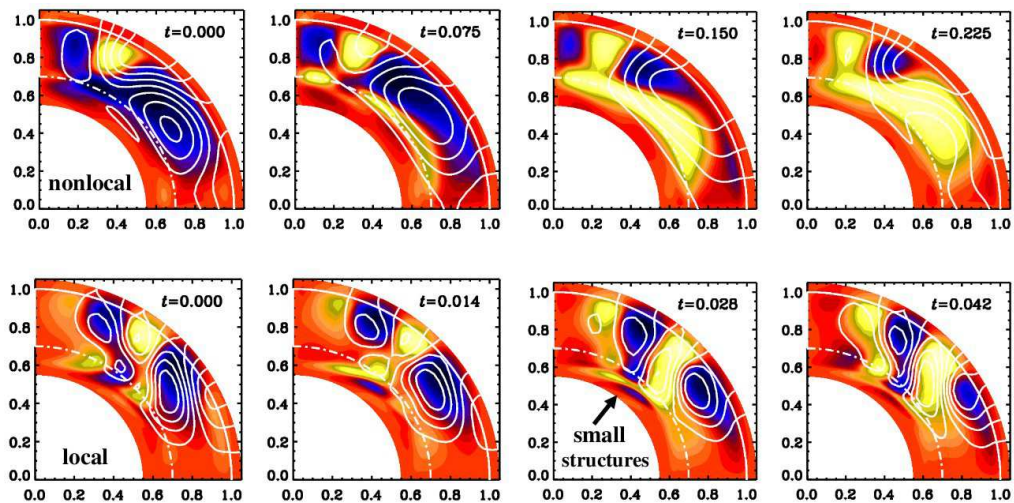}
\end{center}\caption[]{
{\em Top}: Field lines in the meridional plane together with a color-coded
representation of the toroidal field (dark/blue shades indicate
negative values and light/yellow shades positive values).
Evolution of the field structure for model with near-surface shear layer
using the $\partial\meanEMF/\partial t$ equation.
{\em Bottom}: same, but without the $\partial\meanEMF/\partial t$ equation.
The magnetic cycle period is decreased from 0.53 to 0.11 diffusive times
and the excitation conditions enhanced by a factor of five.
Adapted from \cite{BC18}.
}\label{d64_U3e4a}\end{figure}

The beauty of the approach of using \Eq{dEMFdt} lies in the fact that there
is no problem in handling spherical geometry or even nonlinearities
in an {\em ad hoc} manner such as $\alpha$ quenching, as was already
emphasized by \cite{RB12}.
We say here {\em ad hoc}, because the original convolution is linear.

In \Fig{d64_U3e4a} we show a comparison of two models of \cite{BC18}
in spherical geometry with and without spatio-temporal nonlocality.
This model uses solar-like differential rotation contours and turbulent
transport coefficients estimated from mean-field theory.
It shows that spatio-temporal nonlocality implies the absence of small
structures, especially near the lower overshoot layer of the dynamo.
Top and bottom panels cover half a period, so the panels on the right
are similar to those on the left, except for a sign flip.
The cycle period in the model with the $\partial\meanEMF/\partial t$
term included is 0.53 diffusion times, which is about five times longer
than the period of 0.11 of the corresponding conventional models.
For oscillatory solutions such as this one, temporal nonlocality lowers
the excitation conditions of the dynamo, as was already demonstrated
by \cite{RB12}.
In this example, the excitation conditions are lowered by a factor
of about eight.
Below, in \Sec{RflowIII}, we turn to the emergence of a completely new
dynamo effect that occurs just owing to the presence of nonlocality
in time.
Before this, however, we briefly explore the essence of the test-field
method that led to the new insights regarding nonlocality.

\subsection{The test-field method: a way forward}
\label{TFM}

Many of the detailed results discussed below would not have been
discovered without the test-field method.
We therefore briefly review in the following its basic aspects.

Analytic approaches have demonstrated the vast multitude of different
effects, but they are limited in that, for turbulent flows with finite
correlation times, they are only exact at low $\Rm$.
Some methods such as the $\tau$ approach are supposed to work at large $\Rm$, but they
are not rigorous and always subject to numerical verification, using
usually the test-field method.

In essence, the test-field method consists of solving \Eq{inductb}
numerically, subject to given test fields $\meanBB^T$, where the
superscript $T$ denotes one of as many test fields as are needed to
compute uniquely all elements of the $\alpha_{ij}$ and $\eta_{ijk}$
tensors.
In the following, we adopt $xy$ averaging, denoted by an overbar, and
use the two test fields
\EQ
\meanBB^{T_1}=(\cos kz,0,0)\quad\mbox{and}\quad
\meanBB^{T_2}=(\sin kz,0,0).
\EN
For each of them, we find numerically a solution that we call
correspondingly $\bb^{T_1}(\xx,t)$ and $\bb^{T_2}(\xx,t)$.
We then compute the corresponding mean electromotive force
$\meanEMF^{T_1}=\overline{\uu\times\bb^{T_1}}$ and
$\meanEMF^{T_2}=\overline{\uu\times\bb^{T_2}}$.
Inserting this into \Eq{LocalInstantaneous} yields
\EQA
\meanemf_i^{T_1}&=&\meanemf_i^{(0)}+\alpha_{i1}\cos kz-\eta_{i13}\sin kz,\\
\meanemf_i^{T_2}&=&\meanemf_i^{(0)}+\alpha_{i1}\sin kz+\eta_{i13}\cos kz.
\ENA
Here the last index of $\eta_{ijl}$ is $l=3$, because $xy$ averages
only depend on the third spatial coordinate, $z$.
To eliminate $\meanemf_i^{(0)}$, we need solutions for the trivial test
field $\meanBB^{T_0}=\nullvector$.
The solutions $\bb^{T_0}$, and thus $\meanEMF^{(0)}$, may then well be zero,
but there are also cases where they are not---for example if the cross
helicity is finite; see \cite{BR13}.

We are now left with two pairs of unknown coefficients,
$\alpha_{i1}$ and $\eta_{i13}$, for the two nontrivial cases
$i=1$ and $i=2$.
(The third component of $xy$ averaged mean fields is constant because
$\nab\cdot\meanBB=\meanB_{3,3}=0$, so $\meanB_3=0$ if it vanished
initially.)
The two pairs of unknowns are readily obtained by solving a
$2\times2$ matrix problem with the solution \citep{B05QPO}
\EQ
\pmatrix{\alpha_{i1}\cr \eta_{i13}k}=
\pmatrix{~~\cos kz & \sin kz \cr -\sin kz & \cos kz}
\pmatrix{
\meanemf_i^{T_1}-\meanemf_i^{(0)}\cr
\meanemf_i^{T_2}-\meanemf_i^{(0)}},
\label{SolutionVector}
\EN
which yields altogether four coefficients:
$\alpha_{11}$, $\eta_{113}$, $\alpha_{21}$, and $\eta_{213}$.
To get the remaining four coefficients,
$\alpha_{12}$, $\eta_{123}$, $\alpha_{22}$, and $\eta_{223}$,
we need two more test fields, $\meanBB^{T_1}=(0,\cos kz,0)$
and $\meanBB^{T_2}=(0,\sin kz,0)$.
Analogously to \Eq{SolutionVector}, this yields
$(\alpha_{i2},\eta_{i23}k)$ as the corresponding solution vector.

All these coefficients are generally also time-dependent.
For fluctuating fields, as is the case when $\uu$ corresponds
to turbulence, the coefficients are evidently also fluctuating.
This can be relevant for studies of the incoherent $\alpha$-shear
dynamo that will be discussed in \Sec{IncoherentAlphaShear};
see \cite{BRRK08} for such applications.
Another important case is where the test fields themselves are time-dependent.
In fact, this is of immediate relevance to all dynamo problems, where
we expect the mean field to grow exponentially.
Even for a simple turbulent decay problem, $\meanBB$ is time-dependent:
it is exponentially decaying.
Both of these cases were considered by \cite{HB09} using test fields
proportional to $e^{st}$ or $e^{-\ii\omega t}$ with real
coefficients $s$ and $\omega$ that they varied.
This allowed them to assemble the functions $\alpha_{ij}(\omega)$
and $\eta_{ij3}(\omega)$, which led them to the results that for
turbulent flows, both coefficients are, to lowest order, proportional
to $1/(1-\ii\omega\tau)$ with $\tau$ being some relaxation (or
correlation) time, which is proportional to $(\urms\kf)^{-1}$.
The same result was obtained for test fields proportional to $e^{st}$.

\subsection{Alternative approaches to turbulent transport coefficients}
\label{Alternative}

It may be worth noting that there are a few other methods for computing
$\alpha_{ij}$ and $\eta_{ijl}$.
The simplest one is the imposed field method, which is exact in two
dimensions and can then handle also fully nonlinear problems with
magnetic background turbulence \citep{RB10}, as will be discussed in
the next section.
Instead of solving \Eq{inductb}, one solves \Eq{inductB} in the
presence of an imposed field.
It was used to show that $\alpha_{xx}$ and $\alpha_{zz}$ can have
opposite signs in rotating convection \citep{BNPST90}.
In three dimensions, however, turbulence makes the mean field nonuniform,
so the actual electromotive force applies in reality to a problem with
$\alpha$ effect {\em and} turbulent diffusion while using just volume
averages, as if there was no mean current density.
Thus, this method is only of limited usefulness in three dimensions.
A possible way out of this is to reset the fluctuations in regular
intervals \citep{OSB01}.

Another method assumes that, in a time-dependent turbulence
simulation, $\meanEMF$, $\meanBB$, and $\meanJJ$ cover all
possible states, allowing one to obtain all the coefficients
of $\alpha_{ij}$ and $\eta_{ijk}$ after averaging.
This method has even been used to determine spatial nonlocality
\citep{BS02}, but it is not fully reliable, as was later demonstrated
by comparing with the test-field method \citep{B05QPO}.
Nevertheless, some success has been achieved in applications
to accretion disk turbulence \citep{KOMH06} and convection in spherical
shells \citep{Racine,Simard}; see \cite{War18} for a comparative assessment.
Yet another method is multiscale stability theory \citep{LNVW99},
which was recently shown to yield results equivalent to those
of the test-field method \citep{ABNZ15}.

Many of the approaches developed for the induction equation are
also applicable to the momentum equation, where turbulent viscosity,
the anisotropic kinetic alpha (AKA) effect \citep{FSS87}, and the
$\Lambda$ effect \citep{Rue80} are prominent additions.
Turbulent viscosity has been computed by determining the Reynolds
stress in shear flows \citep[e.g.][]{ABL96,SKKL09} or in decay
experiments \citep{YBR03}.
On the other hand, by assuming the turbulent viscosity to
be well approximated by $\nut\approx\urms/3\kf$, it has
also been possible to estimate AKA and $\Lambda$ effects
\citep{PTBNS93,RBMD94,BvR01,KKKBOP15,Kap18}.
However, determining both $\nut$ and $\Lambda$ or AKA effects
at the same time has not yet been successful.

An alternative or extension to mean-field theory in the usual
sense is to solve the time-dependent system of one-point and two-point
correlation functions.
This approach goes by the name Direct Statistical Simulations
\citep{TM13,TM17} and has been applied to two-dimensional turbulent shear
flow problems.
The dimensionality of the two-point correlation function doubles for
those directions over which homogeneity cannot be assumed.
On the other hand, the dynamics of the low order statistics is usually
slower than that of the original equations.
In addition, it is possible to reduce the complexity of the problem by
employing Proper Orthogonal Decomposition \citep{ATM17}.
This approach has not yet been applied to magnetohydrodynamics and the
dynamo problem, but it has the potential of being a strong competitor
in addressing the high Reynolds number dynamics of problems of
astrophysical and geophysical relevance.

\subsection{From quasilinear to fully nonlinear test-field methods}

The test-field equations are readily available in some publicly
available codes, so for example in the {\sc Pencil Code}\footnote{
\url{https://github.com/pencil-code}} \citep{B05QPO} and in {\sc
Nirvana}\footnote{\url{http://www.aip.de/Members/uziegler/nirvana-code/}}
\citep{GZER08,Gre13}.
To newcomers in the field, it is always somewhat surprising that the
test-field equations, i.e., \Eq{inductb} with $\meanBB$ being replaced
by $\meanBB^T$, can be solved without the magnetic field module being
included at all.
The reason is that the turbulent transport coefficients characterize just
properties of the flow.
Thus, the number of equations being solved is just the four or five
hydrodynamic equations (either without or with energy equation included)
together with the four versions of \Eq{inductb} for each of the four
test fields---or more, if more test-fields are needed \citep[see]
[for a case where nine vector equations were solved]{War18}.
However, if the magnetic field module is invoked, the magnetic
field (which is different from the test fields) can grow and backreact
onto the flow.
Thus, one obtains turbulent transport coefficients that are being modified
by the magnetic field.
This method is often referred to as the quasi-kinematic method and has
been used on various occasions to the magnetic quenching of $\alpha$
and $\etat$ \citep{BRRS08,KRBKK14}.
The limits of applicability of this method are still being investigated.
Fully nonlinear approaches have been investigated; see \cite{CHP10}
and \cite{RB10}.
In those approaches, one also solves \Eq{dudt_approx} for the fluctuating
velocity.

The perhaps most striking counter example where the quasi-kinematic
test-field method fails is that of a magnetically forced Roberts flow.
This can easily be seen by computing the $\alpha$ effect with the imposed
field method in two dimensions, i.e., when there is no interference from
turbulent diffusion or other terms.
In such cases, the imposed field and fully nonlinear methods agree,
while the quasi-kinematic method gives even the wrong sign of $\alpha$;
see \cite{RB10} for details.
Magnetically driven flows could in principle be realized by currents
flowing through wires within the flow.
This is a special situation that is not encountered in astrophysics.
However, \cite{RB10} speculated that flows exhibiting small-scale dynamo
action could provide another example where the quasi-kinematic method
fails, but this still needs to be demonstrated.

\subsection{Dynamo effects from memory alone: Roberts flow~III}
\label{RflowIII}

Let us now discuss a remarkable result that has emerged by
applying the test-field method to simple flow fields.
The particular flow field considered here is referred to as
Roberts flow III, which is one of a family of flows he studied
\citep{Rob72}.
In Fourier space, as discussed in \Sec{TheTrick}, the nonlocality
in time corresponds to a division by $1-\ii\omega\tau$.
This leads to an imaginary contribution in the dispersion relation that
can turn a non-dynamo effect into a dynamo effect.
An example is the pumping term, also known as turbulent diamagnetism
\citep{Zel57,Rae69c}.
It corresponds to a contribution to $\meanEMF$ of the form
$\ggamma\times\meanBB$, where $\ggamma$ is a vector that leads to
advection-like transport of the mean magnetic field without actual
material motion.
It corresponds to a transport down the gradient of turbulent intensity.
We return to this aspect in \Sec{DownwardPumping}.
Note also that the $\ggamma$ term corresponds to an off-diagonal
contribution to the $\alpha$ tensor of the form
\EQ
\alpha_{ij}=-\epsilon_{ijk}\gamma_k.
\EN
Quite generally, the $\ggamma$ term implies that the dispersion relation
for the complex growth rate $\lambda(k)$ takes the form
\EQ
\lambda(k)=-\ii\kk\cdot\ggamma-(\eta+\etat)k^2,
\label{GammaDisp}
\EN
where we have ignored other terms such as additional anisotropies,
which do not enter for Roberts flow~III.

Evidently, if we replace $\ggamma\to\ggamma^{(0)}/(1-\ii\omega\tau)$,
neglecting here the $k^2/\kf^2$ term from the spatial nonlocality,
and assuming $\omega\tau\ll1$, then
$-\ii\kk\cdot\ggamma\approx-\ii\kk\cdot\ggamma^{(0)}
+\omega\tau\kk\cdot\ggamma^{(0)}$.
Here, $\omega=\ii\lambda$ is a complex frequency and is used
interchangeably with $\ii\lambda$.
Thus, there can be growth resulting from the second term if
$\omega\tau\kk\cdot\ggamma^{(0)}>\etat k^2$.
Such solutions are always oscillatory and show migratory dynamo waves
in the direction of $\ggamma^{(0)}$.

Solutions of the type discussed above have been found in direct numerical
simulations of Roberts flow~III \citep{RDRB14}.
We now discuss the basic properties of one of their solutions in more
detail.
This flow is given by \citep{Rob72}
\EQ
\uu=u_0\pmatrix{
\quad \sin k_0 x\,\cos k_0 y\cr
     -\cos k_0 x\,\sin k_0 y\cr
\half(\cos2k_0 x+\cos2k_0 y)}
\quad\quad\mbox{(Roberts flow~III)},
\EN
where $u_0$ is an amplitude factor and $k_0$ is the wavenumber of the flow.
Both parameters enter in the definition of the magnetic Reynolds number,
$\Rm=u_0/\eta k_0$.
\cite{RDRB14} found that dynamo action with a mean field proportional
to $\exp[\ii(kz-\omega t)]$ is possible when $k/k_0\la0.78$.
This requires tall domains; in this case with $L_z/L_x=1/0.78$.
In the limit $k\to0$, there is large-scale dynamo action when $\Rm\ga2.9$.
The mean field is oscillatory with a frequency that is at onset about
$\omega\approx0.037\,u_0k_0$.

The marginally excited dynamo solution for Roberts flow~III is already
beyond the validity of SOCA, so the $\uu\times\bb-\overline{\uu\times\bb}$
term in \Eq{UxBprime} cannot be neglected.
In fact, within the limitations of SOCA, which is only valid for small
$\Rm$, no mean-field dynamo can be obtained for Roberts flow~III.
This is because, in the mean-field formalism, the $\gamma$ term was found
to emerge quadratically in $\Rm$, suggesting that it is a higher-order
effect.
\cite{RDRB14} discussed in detail a particular example where $\Rm=6$ and
$k/k_0=0.4$.
The growth rate was found to be $0.047\,u_0k_0$ and the frequency was
$0.29\,u_0k_0$.
In Fourier space, the turbulent magnetic diffusivity kernel was
found to be $\etat(k,\omega)=(0.21+0.03\,\ii)\,u_0/k_0$, which has
only a small imaginary part corresponding to a weak memory effect,
and $\gamma(k,\omega)=(0.73+0.27\,\ii)\,u_0$, which has a significant
imaginary part corresponding to a strong memory effect.
It is this term that is responsible for the positive growth rate.
These complex coefficients match the dispersion relation given by
\Eq{GammaDisp} and reproduce the correct complex growth rate.

Describing spatio-temporal nonlocality with an evolution equation for
$\meanEMF$ is an approximation that is inaccurate for two reasons.
First, in \Eq{EMFomk} there are in general higher powers of $k$
and $\omega$, and second, the $k$ and $\omega$ dependencies of
$\tilde\alpha_{ij}$ and $\tilde\eta_{ijk}$ in \Eq{EMFomk} are
usually not the same; see \cite{RDRB14} for details.
The main point of using such an approximation is to do better than
just neglecting spatio-temporal nonlocality altogether, as is still
done in the vast majority of astrophysical applications.
The differences are substantial, as was already demonstrated in
\Fig{d64_U3e4a}.
We see this again in the present example where the simple
evolution equation \eq{dEMFdt} for $\meanEMF$ reproduces thus a
qualitatively new dynamo effect.

\subsection{Other Roberts flows and generalizations}

In his original paper, \cite{Rob72} discussed altogether four flows.
All the Roberts flows are two-dimensional with the same flow vectors in
the horizontal $(x,y)$ directions, but different $xy$ patterns in the
$z$ direction.
His flow~II is closely related to flow~III discussed above; see
\cite{RDRB14} for details.
It also leads to dynamo waves resulting from the off-diagonal terms
$\alpha_{xy}$ and $\alpha_{yx}$ of the $\alpha$ tensor with dynamo action
owing to the memory term.
The only difference is that here $\alpha_{yx}=\alpha_{xy}$
while for flow~III we had $\alpha_{yx}=-\alpha_{xy}=\gamma$.
Therefore, there are dynamo waves traveling in opposite directions for
$\meanB_x$ and $\meanB_y$.

Another interesting and very different example is Roberts flow~IV,
which is given by
\EQ
\uu=u_0\pmatrix{
\quad \sin k_0 x\,\cos k_0 y\cr
     -\cos k_0 x\,\sin k_0 y\cr
\quad \sin k_0 x            }
\quad\quad\mbox{(Roberts flow~IV)}.
\EN
It also produces large-scale magnetic fields that ``survive'' horizontal
averaging, but in this case the governing dispersion relation is just
of the form
\EQ
\lambda(k)=-[\eta+\etat(k)] k^2,
\EN
where $\etat(k)$ was found to be sufficiently negative for $k\la0.8\,k_0$,
but positive (corresponding to decay) for larger values of $k$ \citep{DBM13}.
Thus, on small length scales, the solution is always stable.

For completeness, let us mention that negative turbulent diffusivities can
also be found for some compressible flows. However, in all those cases
the destabilizing effect is never strong enough to overcome the
microphysical value, i.e., $\etat+\eta$ is still positive \citep{Rae11}.

The most famous Roberts flow is his flow~I, because it is helical and
therefore leads to an $\alpha$ effect.
Moreover, its helicity is maximal with
$\overline{\oo\cdot\uu}=k_0\,u_0^2$.
The flow is given by
\EQ
\uu=u_0\pmatrix{
\quad \sin(k_0 x+\varphi_x)\,\cos(k_0 y+\varphi_y)\cr
     -\cos(k_0 x+\varphi_x)\,\sin(k_0 y+\varphi_y)\cr
\!\!\sqrt{2}\sin(k_0 x+\varphi_x)\,\sin(k_0 y+\varphi_y)}
\;\;\mbox{(Roberts flow~I for $\varphi_x=\varphi_y=0$)},
\label{RflowI}
\EN
where $\varphi_x=\varphi_y=0$ will be assumed at first.
This flow leads to a standard $\alpha$ effect dynamo with a dispersion
relation that is the same as for isotropic turbulence \citep{Mof70},
namely
\EQ
\lambda(k)=\pm|\alpha\kk|-[\eta+\etat]k^2,
\EN
where dynamo action is only possible for the upper sign.
The dynamo is non-oscillatory.
We return to $\alpha$ effect dynamos further below, but before doing
so, let us briefly discuss an interesting feature that arises when
generalizing this flow to the case with time-dependent phases, as done
by \cite{GP92}, who assumed
\EQ
\varphi_x=\epsilon\cos\omega t,\quad\quad
\varphi_y=\epsilon\sin\omega t,
\EN
where $\epsilon$ and $\omega$ are additional parameters characterizing
what is now generally referred to as the Galloway--Proctor flow.
One normally considers a version of this flow that is rotated by
$45\degr$, which allows one to fit two larger cells into the domain
instead of the four cells in \Eq{RflowI}.
This flow is a time-dependent generalization of Roberts flow~I.
This time-dependence is of particular interest in that it allows
the dynamo to become ``fast,$\!\!$'' which means that it can maintain a
finite growth rate in the limit of large magnetic Reynolds numbers,
$\Rm=\urms/\eta\kf\gg1$.

Numerical investigations of the Galloway--Proctor flow revealed the
occurrence of an unexpected pumping effect, i.e., $\ggamma\neq\nullvector$
\citep{CHT06}.
This is because, owing to the circular polarization of this flow,
the symmetry between $\zzz$ and $-\zzz$ is broken \citep{RB09}.
Remarkably, such a $\gamma$ effect does not emerge in the SOCA
approximation which neglects the $\uu\times\bb-\overline{\uu\times\bb}$
term in \Eq{UxBprime}.
Numerical computations of $\gamma$ with the test-field method showed that,
indeed, for $\Rm\to0$, one has $\gamma\to0$.
Furthermore, as $\Rm\to0$, we have $|\gamma|\propto\Rm^5$, which is
a rather steep dependence.
Analogously to the $\gamma$ effect discussed in \Sec{RflowIII}, where
$|\gamma|$ increases quadratically with $\Rm$, this again suggests that
this effect can only be described with a higher-order approximation in
$\Rm$ that is here higher than fourth order.
Indeed, as shown by \cite{RB09}, a fourth-order approximation still does
not capture this effect.

\subsection{Horizontal averaging is not always suitable}
\label{HorizontalAveraging}

Discontent with the use of horizontal averaging was expressed in the
work of \cite{Gent13a,Gent13b}, who used averaging over a Gaussian kernel
as an alternative.
Ultimately, the usefulness of a particular averaging procedure can only
be judged at the end, when we know the answer, what kind of large-scale
field can be generated.
The averaging procedure should be able to capture the expected class of
large-scale fields.
As an example, let us mention here a result of \cite{DBM13}, who did
not find a negative eddy diffusivity dynamo for the Taylor-Green flow.
This was indeed true for horizontal averaging, but not for vertical
($z$) averaging, in which case the mean fields are two-dimensional.
Such solutions were found by \cite{ABNZ15}, who presented several examples
where the field survives $z$ averaging, but not $xy$ averaging.
A related example was found by \cite{BEB16} using shearing box
accretion disk simulations with a shear flow $u_y=Sx$ and $S=\const$.
They reported the emergence of different large-scale fields, depending
on whether they employed $xy$ or $yz$ averaging.

The advantage of any of the averages discussed so far is that they obey
the Reynolds rules.
A practical example is azimuthal averaging in a sphere.
However, such averaging fails to describe nonaxisymmetric mean fields.
Alternative averaging procedures such as spatial filtering are
problematic in that they do not obey the Reynolds averaging rules;
see \cite{Rae95,Rae14}.
Ensemble averaging obeys the Reynolds rules and could describe
nonaxisymmetric mean fields, but the practical meaning of such
averaging is unclear \citep{Hoyng03}.
As will be discussed in \Sec{StellarSurfaceStructures} in more detail,
rapidly rotating stars with weak {\em differential} rotation are likely
to exhibit nonaxisymmetric mean fields.
In that regime, the excitation conditions of such dynamos with an
azimuthal order of $m=1$ are comparable to those of axisymmetric
dynamos \citep{Rae80,Rae86a}.
Such solutions are now commonly found for rapidly rotating stars;
see \cite{Viviani18} for recent simulations.

\subsection{Quenching of $\alpha$: self-inflicted anisotropy}

As the magnetic field grows and its energy density becomes comparable to
the kinetic energy density, the Lorentz force in the momentum equation
begins to become important.
This tends to decrease $\alpha$ and $\etat$ in such a way as to saturate
the dynamo.
Assuming that our mean fields correspond to just planar averaging over
the periodic $x$ and $y$ directions, they no longer depend on $x$ and $y$.
It is therefore clear that $\meanBB$ is just a function of $z$ and $t$.
Moreover, since $0=\nab\cdot\meanBB=\meanB_{z,z}$, we have
$\meanB_z=\const$ and, unless $\meanB_z$ is initially finite,
it must vanish at all later times.
For a dynamo driven essentially by an $\alpha$ effect, the $\meanBB$ with
only $x$ and $y$ components must be an eigenfunction of the curl operator.
This applies to all dynamos driven by a helical flow, such as the laminar
Roberts flow~I, and also to three-dimensional helical turbulence, for example.
In a periodic domain $0<z<L_z$, the eigenfunction is given by
\EQ
\meanBB=\pmatrix{ \sin(k_1 z+\varphi)\cr \cos(k_1 z+\varphi)\cr 0},
\label{EigenPos}
\EN
where $k_1=\pm2\pi/L_z$ is the smallest wavenumber of the field in the
$z$ direction and $\varphi$ is an arbitrary phase which is only determined
by the initial conditions.
Note that $\nab\times\meanBB=k_1\meanBB$, so $\meanBB$ is indeed an
eigenfunction of the curl operator.
The eigenvalue $k_1$ is positive (negative) if $\alpha$ is positive
(negative).

Once the magnetic field reaches equipartition strength with the flow,
which we now assume to be driven by a forcing term in the momentum
equation, the magnetic field saturates owing to the action of the Lorentz
force in this momentum equation.
The resulting changes to the flow begin to affect the $\alpha$
tensor, which then inevitably attains an anisotropy proportional to
$\meanB_i\meanB_j/\meanBB^2$ \citep{Rob93}.
We call this self-inflicted anisotropy.
Thus, even if the $\alpha$ tensor was initially isotropic (which is here
the case in the $xy$ plane), it would become anisotropic at saturation
and is then of the form
\EQ
\aalpha=\alpha_0(\meanB)\,\pmatrix{1&0&0\cr0&1&0\cr0&0&0}-\alpha_1(\meanB)\,
\pmatrix{\sin^2 k_1 z & \sin k_1 z \cos k_1 z&0\cr
\sin k_1 z\cos k_1 z & \cos^2 k_1 z&0\cr0&0&0},
\label{NonlinSat}
\EN
where we have assumed $\varphi=0$ for simplicity and
$\meanB\equiv|\meanBB|$ because of $\sin^2 k_1 z+\cos^2 k_1 z=1$,
so the anisotropy is no longer apparent.
Note that $\alpha\meanBB=(\alpha_0-\alpha_1)\meanBB$.
This form of $\aalpha$ with $\alpha_1(\meanB)$ having the opposite
sign of $\alpha_0(\meanB)$ was confirmed by numerical simulations
using the test-field method \citep{BRRS08}.
Certain aspects of it were also verified with the imposed field
method where one neglects the $\eeta$ tensor and simply measures
$\meanEMF=\bra{\uu\times\bb}$ in a simulation and computes then
$\alpha_{ij}$ from $\meanemf_i/\meanB_j$ \citep{HDSKB09}.

\subsection{An insightful experiment with an independent induction equation}

\cite{CT09} were the first to study the nature of solutions to
an independent induction equation,
\EQ
{\partial\ZZ\over\partial t}=\nab\times\left(\UU\times\ZZ
-\eta\nab\times\ZZ\right),
\label{inductZ}
\EN
with a new vector field $\ZZ$ instead of $\BB$, but with the same quenched
velocity field $\UU(\BB)$, which is the solution to the momentum equation
with the usual Lorentz force $\JJ\times\BB$.
The result was surprising in that the dynamo did not saturate by
``relaxing the system to a state close to marginality or by suppressing
the chaotic stretching in the flow'' \citep{CT09}.
They argued further ``that this process is very subtle and not in concord
with any of the previously suggested theories.''
Indeed, the naive expectation would be $\ZZ\propto\BB$, i.e., a field
proportional to the one that led to the now saturated dynamo, whose flow
we used in \Eq{inductZ}.
However, the growth rate of such a $\ZZ$ would be exactly zero.
In other words, we have
\EQ
\alpha\meanZZ=\alpha_0\meanZZ,\quad\mbox{while}\quad
\alpha\meanBB=(\alpha_0-\alpha_1)\meanBB.
\EN
Thus, if there were another solution that could actually grow under the
influence of the velocity field $\UU(\BB)$, it would be the more preferred
solution to \Eq{inductZ}.
Given that $\UU(\BB)$ is helical, we expect nontrivial horizontally
averaged fields $\meanZZ$ to be a solution of the associated mean-field
problem of \Eq{inductZ}, but with an $\alpha$ tensor given still by
\Eq{NonlinSat}, i.e., with $\meanBB$ rather than $\meanZZ$.
Given that $\alpha_1(\meanB)$ and $\alpha_0(\meanB)$ have opposite signs,
an essential contribution to the quenching comes from the second term.
Therefore, solutions $\meanZZ$ that belong to the nullspace of the
matrix $\meanB_i\meanB_j$ would not be quenched by this term.
This is indeed what \cite{TB08} found; their $\meanZZ$
was a $90\degr$ phase-shifted version of $\meanBB$, i.e.,
$\meanZZ(z)=\meanBB(z+\pi/2k_1)$.
Indeed,
\EQ
\pmatrix{\sin^2 k_1 z & \sin k_1 z \cos k_1 z&0\cr
\sin k_1 z\cos k_1 z & \cos^2 k_1 z&0\cr0&0&0}
\pmatrix{ \cos k_1 z \cr-\sin k_1 z\cr 0}=\nullvector,
\EN
so this $\meanZZ$ is not being quenched by this second term in
\Eq{NonlinSat}.
Thus, $|\meanZZ|$ continues to grow exponentially.
Some quenching might still occur because of a change of $\alpha_0(\meanB)$,
but in the experiments of \cite{TB08}, this effect was small.
This remarkable, but perfectly understandable behavior in the evolution
of $|\meanZZ|$ provides another independent verification of the quenching
expression given by \Eq{NonlinSat}.

\subsection{Catastrophic quenching}
\label{CatastrophicQuenching}

Early work with the imposed field method using a uniform magnetic field
$\BB_0=\const$ resulted in an $\alpha$ effect whose value seemed to be
quenched in an $\Rm$-dependent fashion.
\cite{BF00} called this {\em catastrophic} quenching, because $\alpha$
would be catastrophically small in the astrophysically relevant case of
large $\Rm$.
This was first suggested by \cite{VC92} and confirmed numerically by
\cite{CH96}.
This result irritated the astrophysics community for some time.
Indeed, it seemed a bit like a crisis to all of mean-field theory and,
maybe, we would not have had this special edition of the Journal of
Plasma Physics (JPP) if this quenching was really as catastrophic as it
seemed at the time!

The solution to the catastrophic quenching problem turned out to be
another highlight of dynamo theory and has its roots in an early finding
by \cite{PFL76}.
They realized that, in the nonlinear case at sufficiently large $\Rm$,
the $\alpha$ effect has a new contribution which is not just proportional
to the mean kinetic helicity density $\overline{\oo\cdot\uu}$, as
stated in the beginning in \Eq{alpha0Formula}, but there is a term
proportional to the mean current helicity density from the fluctuating
fields $\overline{\jj\cdot\bb}$, where $\jj=\nab\times\bb/\mu_0$ is the
small-scale current density.
This term emerges naturally from the $\overline{\dot{\uu}\times\bb}$
term in \Eq{TwoTerms} when using the $\tau$ approximation; see
\Sec{TauApprox}.
Thus, we have \citep{PFL76}
\EQ
\alpha_0=-\onethird\tau\left(\overline{\oo\cdot\uu}
-\overline{\jj\cdot\bb}/\meanrho\right),
\label{alphaJBFormula}
\EN
where $\meanrho$ is the mean fluid density.
However, if the small-scale magnetic field is still
approximately statistically isotropic, the small-scale current helicity,
$\overline{\jj\cdot\bb}$, must be approximately
$\kf^2\,\overline{\aaaa\cdot\bb}/\mu_0$, where $\aaaa$ is the magnetic
vector potential of the small-scale field, $\bb=\nab\times\aaaa$.
Interestingly, $\overline{\aaaa\cdot\bb}$ is constrained, on the one
hand, by $\overline{\AAA\cdot\BB}$, i.e., the mean magnetic helicity
density of the total field, which obeys a conservation equation, and
on the other hand by $\meanAA\cdot\meanBB$, which is the result of the
mean-field dynamo problem \citep{HB12}, i.e.,
\EQ
{\partial\over\partial t}\meanAA\cdot\meanBB=2\meanEMF\cdot\meanBB
-2\eta\mu_0\meanJJ\cdot\meanBB-\nab\cdot(\meanFF_{\rm m}-\meanEMF\times\meanAA),
\label{dAmBmdt}
\EN
where $\meanFF_{\rm m}$ is the magnetic helicity flux from the large-scale
field and $\meanEE=\eta\mu_0\meanJJ-\meanUU\times\meanBB$ is the mean
electric field without the $\meanEMF$ term.
Thus, $\overline{\aaaa\cdot\bb}$ must obey the equation \citep{KR82,KMRS00}
\EQ
{\partial\over\partial t}\overline{\aaaa\cdot\bb}=-2\meanEMF\cdot\meanBB
-2\eta\mu_0\overline{\jj\cdot\bb}-\nab\cdot(\meanFF_{\rm f}+\meanEMF\times\meanAA),
\label{dabdt}
\EN
so that the sum of \Eqs{dAmBmdt}{dabdt} is equal to
\EQ
{\partial\over\partial t}\overline{\AAA\cdot\BB}=
-2\eta\mu_0\overline{\JJ\cdot\BB}-\nab\cdot\meanFF_{\rm tot},
\label{dABdt}
\EN
where $\meanFF_{\rm tot}=\meanFF_{\rm m}+\meanFF_{\rm f}$ is the sum
of magnetic helicity fluxes from the mean and fluctuating fields,
respectively.
\EEq{dabdt} can easily be formulated as an evolution equation for
$\alpha$, or at least its magnetic contribution, as was first done
by \cite{KR82}.

A few additional comments are here in order.
First, analogous to the pair of terms $\pm2\meanEMF\cdot\meanBB$ in
\Eqs{dAmBmdt}{dabdt}, we have isolated the pair $\mp\meanEMF\times\meanAA$
underneath the corresponding flux divergence terms.
This was first done by \cite{HB12}, who found these to give important
contributions, especially to the flux in the equation for the small-scale
magnetic helicity.
This term complements a corresponding term of opposite sign in the
equation for the large-scale magnetic helicity, but it does not contribute
to the total magnetic helicity flux.
Second, it can be advantageous to solve directly the equation for the
total magnetic helicity flux, as done by \cite{HB12}.
This ensures that mutually canceling terms do not contribute
``accidently'' (as a result of inaccurate approximations) to the total
magnetic helicity flux.
This approach has been adopted by \cite{Pip13,Pip13b} and \cite{PK13,PK16}
to model the solar dynamo; see also \cite{Pip15,Pip17}.

When formulated as an evolution equation for $\alpha$, the approach
described above is referred to as ``dynamical'' quenching.
This is not an alternative to the ``algebraic'' quenching, which
describes the functional dependencies of $\alpha_0(\meanB)$ and
$\alpha_1(\meanB)$ in \Eq{NonlinSat}, but it is an additional contribution
to $\alpha_0(\meanB)$, and has in principle also additional anisotropic
contributions \citep{RK07,Pip08}.
It provides a feedback from the growing or evolving $\meanAA\cdot\meanBB$
that is necessary to obey the total magnetic helicity \Eq{dABdt}.

As pointed out by \cite{RR07}, dynamical quenching has not
been derived rigorously within mean-field theory, and must rather
be regarded as a heuristic approach.
Dynamical quenching does not emerge in the traditional approach
of solving for the fluctuations.
One should expect that the magnetic helicity equation would automatically
be obeyed if one solved the equations for the fluctuations by avoiding
questionable approximations.
At present, however, dynamical quenching is the only known approach that
describes correctly the resistively slow saturation of $\alpha^2$ dynamos
in triply-periodic domains \citep{FB02,BB02,Sub02} found by \cite{Bra01},
as will be discussed in \Sec{ResistivelySlow}.

\begin{figure}\begin{center}
\includegraphics[width=.95\columnwidth]{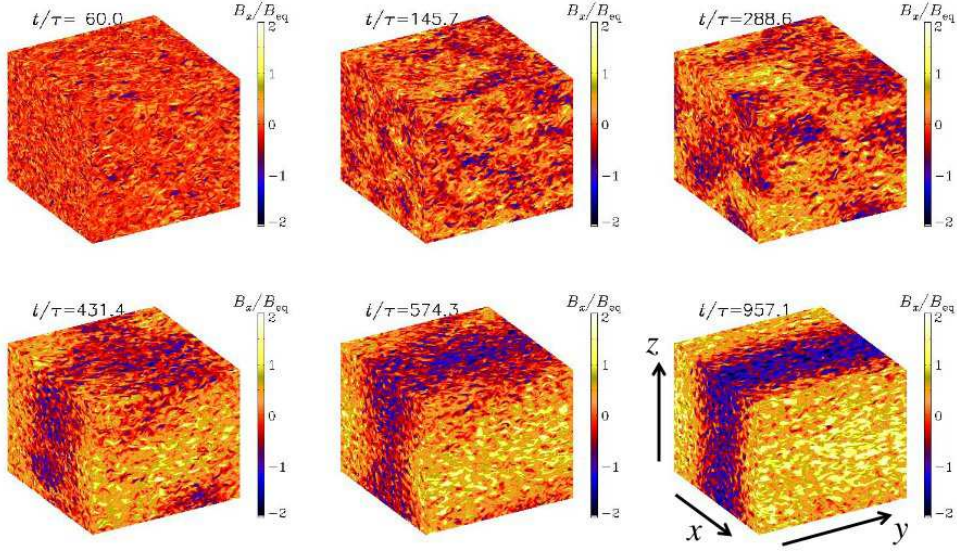}
\end{center}\caption[]{
Visualizations of $B_x/\Beq$ on the periphery of the domain at six
times during the late saturation stage of the dynamo when a large-scale
field is gradually building up.
The small-scale field has reached its final value after $t/\tau\approx100$
turnover times.
The diffusive time is here about 7000 times the turnover time.
The maximum field strength is about twice $B_{\rm eq}$.
}\label{B}\end{figure}

The aforementioned simulations were done with helically forced turbulence,
which led, at late times, to the development of a large-scale magnetic
field of Beltrami type; see \Eq{EigenPos} for one such example, where
the wavevector of the mean field points in the $z$ direction.
In \Fig{B} we show an example of the gradual approach to such a Beltrami
field, which has here a wavevector pointing in the $x$ direction.

The evolution equation for $\alpha$ can also be written in implicit
form with the time derivative of $\alpha$ on the right-hand side as
\citep{Bra08}
\EQ
\alpha={\alpha_{\rm K}+\Rm\left[
\etat{\mu_0\meanJJ\cdot\meanBB/\Beq^2-(\nab\cdot\meanFF_{\rm f})/(2\Beq^2)}
-(\partial\alpha/\partial t)/(2\etat\kf^2)
\right]\over1+\Rm\meanBB^2/\Beq^2},
\label{alp_imp}
\EN
where $\alpha_{\rm K}$ is the $\alpha$ effect in the kinematic limit.
The formulation in \Eq{alp_imp} confirms first of all the early
catastrophic quenching result of \cite{VC92} for {\em volume-averaged}
mean fields,
because those are independent of the spatial coordinates and, therefore,
$\mu_0\meanJJ=\nab\times\meanBB=\nullvector$.
The periodicity implies $\nab\cdot\meanFF_{\rm f}=0$.
Also, they considered a stationary state, so $\partial\alpha/\partial t=0$.
Thus, all the factors of $\Rm$ in the numerator vanish and therefore
we have $\alpha=\alpha_{\rm K}/(1+\Rm\meanBB^2/\Beq^2)$, as predicted
by \cite{VC92}.
In general, however, the presence of any of the three additional terms in
the numerator multiply $\Rm$ and are therefore of the same order as those
in the denominator.
This should readily alleviate the threat of an $\Rm$-dependent quenching.
Interestingly, \Eq{alp_imp} applies also when the dynamo is not
driven by the $\alpha_{\rm K}$ term, but by the shear-current effect,
for example \citep{BS05b}.
Thus, somewhat paradoxically, we could say that an $\alpha$ effect can
be quenched even if there is no $\alpha$ to begin with.

In the absence of magnetic helicity fluxes, i.e., when
$\nab\cdot\meanFF_{\rm f}=0$, as in the present case of homogeneous
turbulence with periodic boundary conditions, the time evolution is
inevitably controlled by a resistively slow term.
This somewhat surprising constraint for {\em homogeneous} helical
turbulence can be understood quite generally---even without resorting
to any mean-field theory, i.e., without talking about $\alpha$ effect
and turbulent magnetic diffusivity.
This will be discussed next.

\subsection{Resistively slow saturation in homogeneous turbulence}
\label{ResistivelySlow}

To describe the late saturation phase, we invoke the magnetic
helicity equation for the whole volume, which is assumed to be
either periodic or embedded in a perfect conductor.
Volume averages will be denoted by angle brackets.
Thus, we have
\EQ
{\dd\over\dd t}\bra{\AAA\cdot\BB}=-2\eta\mu_0\bra{\JJ\cdot\BB},
\label{dABvoldt}
\EN
which is the same as \Eq{dABdt}, but without the magnetic helicity
flux divergence term.
(For the volume averages employed here, this would lead to a surface term,
which vanishes for periodic or perfectly conducting boundaries.)
This equation highlights an important result for the steady state, namely
\EQ
\bra{\JJ\cdot\BB}=0\quad\quad
\mbox{(for any steady state in triply periodic domains)}.
\label{JB0}
\EN
This sounds somewhat boring, but becomes immediately interesting when
realizing that mean fields and fluctuations can both be finite, i.e.,
\EQ
\bra{\jj\cdot\bb}=-\bra{\meanJJ\cdot\meanBB}\neq0,
\label{jb0}
\EN
so that $\bra{\JJ\cdot\BB}=\bra{\meanJJ\cdot\meanBB}+\bra{\jj\cdot\bb}=0$,
as required.

\begin{figure}\begin{center}
\includegraphics[width=\columnwidth]{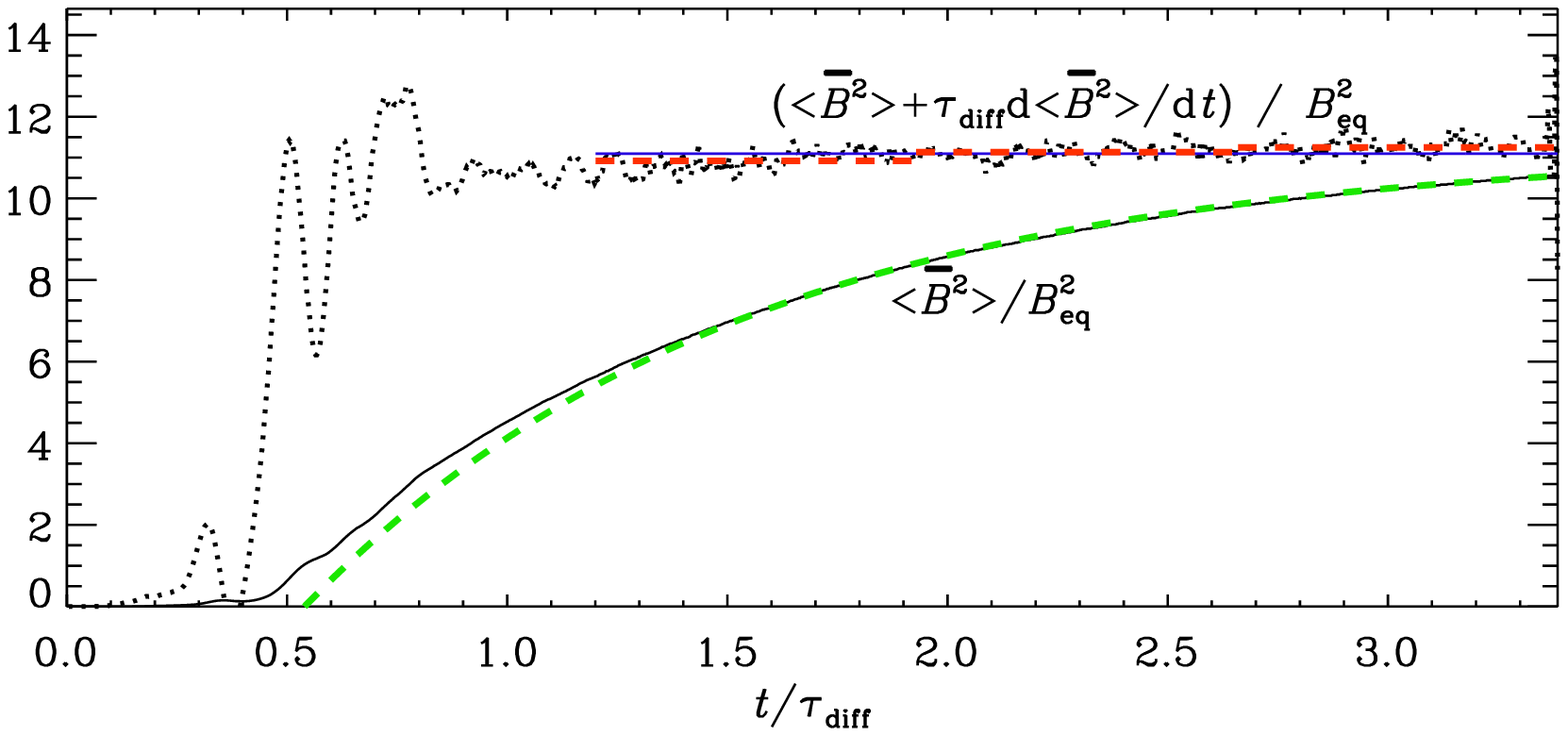}
\end{center}\caption[]{
Evolution of the normalized $\bra{\meanBB^2}$
and that of $\bra{\meanBB^2}+\tau_{\rm diff}\dd\bra{\meanBB^2}/\dd t$ (dotted),
compared with its average in the interval $1.2\leq t/\tau_{\rm diff}\leq3.5$
(horizontal blue solid line), as well as averages over three subintervals
(horizontal red dashed lines).
The green dashed line corresponds to \Eq{B2vol} with
$t_{\rm sat}/\tau_{\rm diff}=0.54$.
}\label{psat}\end{figure}

To describe the gradual approach to the stationary state given by \Eq{JB0},
we have to retain the time derivative in \Eq{dABvoldt}.
Writing $\bra{\AAA\cdot\BB}=\bra{\meanAA\cdot\meanBB}+\bra{\aaaa\cdot\bb}$,
and assuming that, in the late saturation phase, the quadratic correlations
of the fluctuations are already constant and only the correlations of mean
fields are not, we can omit the time derivative of $\bra{\aaaa\cdot\bb}$.
Furthermore, we assume magnetic fields with positive (negative) magnetic
helicity at small scales, i.e.,
\EQ
\mu_0\bra{\jj\cdot\bb}\approx\pm\kf\bra{\bb^2}\approx\kf^2\bra{\aaaa\cdot\bb},
\label{jbk2ab}
\EN
and that $\bra{\bb^2}\approx\mu_0\bra{\rho\uu^2}\equiv\Beq^2$, which is
the square of the equipartition value.
Here, the upper (lower) signs refer to positive (negative) magnetic
helicity at small scales.
Furthermore, owing to \Eq{EigenPos}, we have
\EQ
\meanJJ\cdot\meanBB=\mp k_1\meanBB^2=k_1^2\,\meanAA\cdot\meanBB,
\EN
which is, for pure modes with wavenumber $k_1$, constant in space.
However, this relation is no longer exact for a superposition of modes.
Thus, with these provisions, \Eq{dABvoldt} becomes \citep{Bra01}
\EQ
{\dd\over\dd t}\bra{\meanBB^2}=2\eta k_1\kf\Beq^2-2\eta k_1^2\bra{\meanBB^2},
\label{dB2voldt}
\EN
with the solution
\EQ
\bra{\meanBB^2}=\Beq^2{\kf\over k_1}
\left[1-e^{-2\eta k_1^2(t-t_{\rm sat})}\right].
\label{B2vol}
\EN
This agrees with the slow saturation behavior seen first in the simulations
of \cite{Bra01}; see \Fig{psat}.
Here $t_{\rm sat}$ is the time when the slow saturation phase commences;
see the crossing of the green dashed line with the abscissa.
Interestingly, instead of waiting until full saturation is accomplished,
one can obtain the saturation value already much earlier simply by
differentiating the simulation data to compute \citep{CB13}
\EQ
B_{\rm sat}^2\approx
\bra{\meanBB^2}+\tau_{\rm diff}{\dd\over\dd t}\bra{\meanBB^2}.
\EN
Note that the inverse time constant $\tau_{\rm diff}^{-1}=2\eta k_1^2$
in the exponent of \Eq{B2vol} is fixed by the microphysics and does not
involve the turbulent magnetic diffusivity.
This is therefore still in some sense catastrophic, so real astrophysical
dynamos do not work like this, and this is because of magnetic helicity
fluxes.
To demonstrate this in a really convincing way requires simulations at
magnetic Reynolds numbers well in excess of 1000 \citep{DSGB13}.
We discuss magnetic helicity fluxes next.

\subsection{Magnetic helicity fluxes}
\label{MagneticHelicityFluxes}

The most important contribution to the magnetic helicity flux is a
turbulent diffusive flux proportional to the negative gradient of the
magnetic helicity density \citep{HB10}, i.e.,
\EQ
\meanFF_{\rm f}=-\kappa_h\nab\overline{\aaaa\cdot\bb}.
\EN
Such a formulation raises immediately the question of the gauge dependence
of magnetic helicity.
This turns out to be less of an issue than originally anticipated.
A first step in this realization comes from the work of \cite{SB06},
who showed that the magnetic helicity density can be expressed in terms of
a density of linkages, provided the correlation scale is much smaller
than the mean field or system scale.
In reality, of course, a broad range of length scales will be excited,
and this can be described by the (shell-integrated) magnetic helicity
spectrum, $\HM(k)$, which is normalized such that
$\int\HM(k)\,\dd k=\bra{\AAA\cdot\BB}$.
For a general review on astrophysical turbulence discussing also spectra
such as these, see \cite{BN11}.

Magnetic helicity spectra have been obtained from solar observations
\citep{ZBS14,ZBS16,BPS17} and even for the solar wind \citep{MG82,BSBG11},
as will be discussed below.
Such spectra are automatically gauge-invariant owing to the implicit
assumption that, by taking a Fourier transform, one assumes a periodic
domain.
Clearly, this is unrealistic on the largest scales, but this only affects
the magnetic helicity spectra at the smallest wavenumbers.
At all other wavenumbers, the spectrum should be a physically meaningful
quantity and the same in any gauge.

\begin{figure}\begin{center}
\includegraphics[width=\columnwidth]{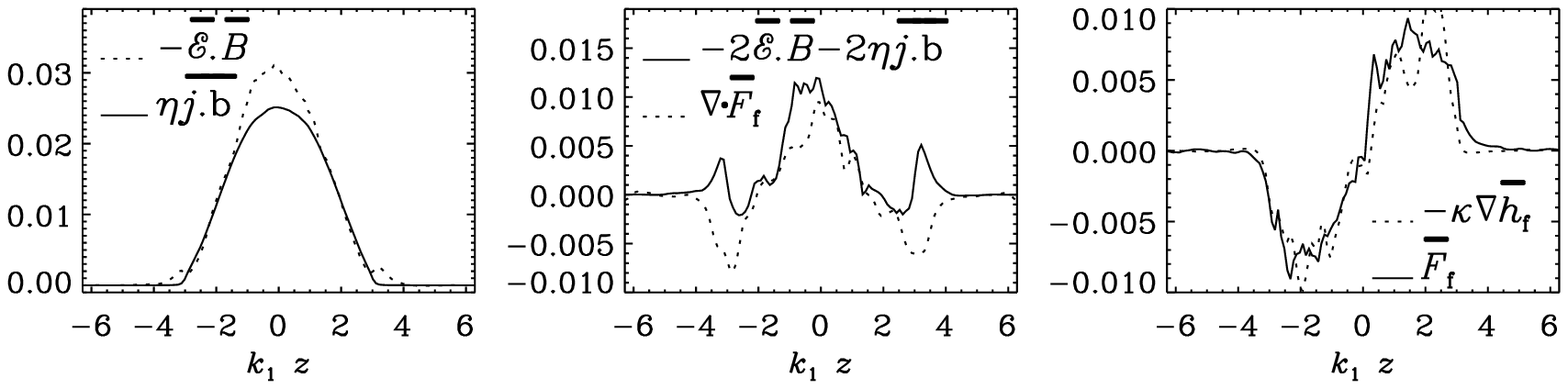}
\end{center}\caption[]{
Time-averaged profiles of $\bra{\meanEMF\cdot\meanBB}$ and
$\eta\bra{\jj\cdot\bb}$ (left panel),
the difference between these terms compared with the magnetic helicity flux
divergence of small-scale fields $\bra{\nab\cdot\meanFF_{\rm f}}$
(middle panel), and the flux itself compared with the Fickian diffusion
ansatz (right-hand panel).
The fluxes are in given in units of $\etatz\Beq^2$ and the
flux divergence is given in units of $k_1\etatz\Beq^2$.
}\label{pflux_profile}\end{figure}

Measurements of magnetic helicity fluxes have been performed by
\cite{HB10} for an $\alpha^2$ dynamo embedded in a poorly conducting halo
and by \cite{DSGB13} for a dynamo with a wind
so one can compare turbulent--diffusive and advective fluxes.
\cite{MCCTB10} have explicitly demonstrated the gauge independence of the
small-scale magnetic helicity flux by working in three different gauges.
In all those cases, it was found that the magnetic helicity flux
divergence is comparable to the Spitzer magnetic helicity production,
$2\eta\mu_0\overline{\jj\cdot\bb}$.
In \Fig{pflux_profile}, we show time-averaged profiles of
$2\meanEMF\cdot\meanBB$ and $2\eta\overline{\jj\cdot\bb}$, as well as the
difference between these two terms compared with the magnetic helicity
flux divergence of small-scale fields, $\nab\cdot\meanFF_{\rm f}$,
and the flux itself compared with the Fickian diffusion ansatz for the
model of \cite{HB10} at $\Rm\approx270$.
We see that the magnetic helicity flux
divergence of small-scale fields is still less than the magnetic helicity
production by the mean electromotive.
Thus, the magnetic helicity flux divergence is still subdominant.
It can therefore not yet alleviate the resistively slow saturation
of the dynamo.
One might hope that this will change at larger values of $\Rm$.
As of now, however, it has not yet been possible to demonstrate this
convincingly.

Most of the dynamo simulations to date are not yet in the asymptotic regime
where $\Rm$ is large enough to alleviate resistively slow saturation.
It would be important to demonstrate more thoroughly to what extent
those dynamos are in the asymptotic regime, and that
\EQ
|\nab\cdot\meanFF_{\rm f}|\approx|2\meanEMF\cdot\meanBB|
\gg|2\eta\mu_0\overline{\jj\cdot\bb}|,
\EN
as one should expect.
Let us emphasize here that, unlike the flux divergence
$\nab\cdot\meanFF_{\rm f}$, the actual helicity fluxes can always be
gauged such that they vanish across an impenetrable boundary by adopting
the gauge $\UU\cdot\AAA=0$ \citep{CHBM11}.
In that case, the magnetic helicity density evolves just like a
passive scaler, i.e.,
\EQ
{\partial\over\partial t}\AAA\cdot\BB=-\nab\cdot[(\AAA\cdot\BB)\,\UU],
\EN
where the flux contribution $(\UU\cdot\AAA)\,\BB$ vanishes; see \cite{HB11}.

\subsection{Oscillatory $\alpha^2$ dynamo: an exactly solvable model
for continued investigations}

Much of the work on catastrophic quenching and resistively slow
saturation has come from studies in periodic domains, where no helicity
fluxes are possible.
To go beyond this limitation, we need to focus on inhomogeneous
conditions and possibly also inhomogeneous turbulence.
A particularly simple system that has not yet been studied in this regard
is the $\alpha^2$ dynamo between a perfectly conducting boundary on
one side ($A_x=A_y=A_{z,z}=0$ in the Weyl gauge) and a vertical field
condition ($A_{x,z}=A_{y,z}=A_z=0$) on the other.

In the following, we discuss a mean-field dynamo with a mean
magnetic vector potential given by $\meanAA=(\meanA_x,\meanA_y,0)$ and
the same boundary conditions, namely $\meanA_x=\meanA_y=0$ on one side
and $\meanA_{x,z}=\meanA_{y,z}=0$ on the other.
Such dynamos have oscillatory solutions that can be written in closed
form as \citep{Bra17}
\EQ
{\cal A}(z,t)\equiv \meanA_x+\ii \meanA_y=A_0\,
\left(e^{\ii k_+z}-e^{\ii k_-z}\right)e^{-\ii\omega t},
\label{calAeqn}
\EN
where the wavenumbers $k_+$ and $k_-$ are complex so as to satisfy
the vacuum boundary condition $\partial{\cal A}/\partial z=0$ on $k_0
z=\pi/2$, with $k_0$ being the lowest wavenumber of the decay mode in
this model, and $A_0$ is an amplitude factor.
The two wavenumbers obey the constraint relation $(k_++k_-)\etaT+\alpha=0$
with $\etaT$ being the total (turbulent plus microphysical) magnetic
diffusivity, and are given by
\begin{equation}
k_+/k_0\approx0.10161896-0.51915398\,\ii,
\end{equation}
\begin{equation}
k_-/k_0\approx-2.6522693+0.51915398\,\ii.
\end{equation}
at the first critical complex eigenvalue defined by the marginal value
of $\alpha$ and the frequency $\omega$ with
\begin{equation}
\alpha k_0+\ii\omega\approx(2.5506504-1.4296921\,\ii)\,\etaT k_0^2.
\label{eigenval}
\end{equation}
\EEq{calAeqn} automatically obeys the perfect conductor boundary condition
${\cal A}=0$ at $z=0$.
These solutions display dynamo waves traveling away from the perfect
conductor boundary toward the vacuum boundary.
This is reminiscent of the work of \cite{Par71}, who found that for
oscillatory $\alpha\Omega$ dynamos, boundary conditions can introduce
behaviors that are not obtained for infinite domains.
Subsequently, \cite{WKTP97} and \cite{TPK97} found that the antisymmetry
condition at the equator plays the role of an absorbing
boundary that led to localized wall modes.
Later, \cite{TPK98} showed that boundary conditions can play a decisive role
in determining the migration direction of traveling waves.

Oscillatory $\alpha^2$ dynamos have been studied numerically in strongly
stratified domains \citep{JBKR16}, but the question of magnetic helicity
fluxes has not yet been addressed.
A model with these boundary conditions, but applied to three-dimensional
turbulence, may be an ideal target to re-address the question of magnetic
helicity fluxes.
This model would be an improvement over previous studies where the
vertical field boundary condition has been used on both ends of the
domain; see \cite{GD94,GD95,GD96} and \cite{BD01}.

A particularly simple mean-field model with nontrivial helicity fluxes
was presented by \cite{BCC09} for a variant of the model presented above.
It revealed for the first time that the magnetic helicity density in
the outer parts of the domain, i.e., in the halo, is reversed.
Its significance was not fully appreciated until later when it was
actually observed in the solar wind \citep{BSBG11}.
Before going into details, let us first discuss what is known about
magnetic helicity in the Sun.

\subsection{$\alpha\Omega$ dynamos}

An important class of dynamos is the $\alpha\Omega$ dynamo.
In addition to the $\alpha$ effect, there is shear or differential rotation,
referred to as $\Omega$ effect.
The dispersion relation of such dynamos has been known since the work
of \cite{Par55}.
In the absence of boundaries, it predicts planar dynamo waves traveling
in the spanwise directions.
For example, in a linear shear flow with $U_y(x)=Sx$ and $S=\const$,
dynamo waves travel in the positive (negative) $z$ direction if the sign
of the product $\alpha S$ is positive (negative).
This has been confirmed in direct numerical simulations \citep{BBS01,KB09}.
In the presence of boundaries in the $z$ direction, as is the case in certain
convection setups, the dynamo can become nonoscillatory.
This was also confirmed in simulations \citep{KKB08,HP09}.
We will return to this subject on several occasions, because such
dynamos are believed to play important roles in the solar dynamo
(\Sec{DynamoDilemma}), stellar dynamos (\Sec{AntiquenchedStellarDynamos}),
and accretion disk dynamos (\Sec{IdentifyingAlphaOmega}).

\section{The solar dynamo}

The measurement of solar magnetic helicity has always been concerned
with the gauge dependence and topological nature of magnetic helicity.
This led to the development of the relative magnetic helicity
\citep{BF84,FA85}, a gauge-invariant formulation of the magnetic helicity
in a given open domain obtained by making reference to a potential field
obeying the same boundary conditions on the periphery of the domain.
In the following, however, we focus on magnetic helicity spectra
and discuss their significance and advantages over the full volume
integrated quantity.

\subsection{Magnetic helicity spectra}

It has long been speculated that astrophysical dynamos might be in some
way magnetically driven, i.e., driven by a magnetic instability such as
the magneto-buoyancy \citep{HP88} or magneto-rotational instabilities
\citep{BH98}.
This motivated the study of dynamos with a forcing term in the
induction equation, as was first done by \cite{PFL76}.
Although this reasoning may not apply in practice, such models do have
the interesting property that they have the same sign of magnetic helicity
at all length scales \citep{PB12b}.
By contrast, kinetically driven dynamos result in a bihelical spectrum
with opposite signs of magnetic helicity at large and small length scales
\citep{Bra01,BB03}.
Thus, to distinguish between these rather different scenarios, we need
to compute the spectrum of magnetic helicity.
In particular, we must look for the possibility of different
signs of magnetic helicity at different scales or wavenumbers.
It is therefore not enough to obtain the magnetic helicity of the total
field, $\bra{\AAA\cdot\BB}=\int\HM(k)\,\dd k$, but the detailed scale
dependence through $\HM(k)$.  For a particular active region on the solar
surface, AR~11158, the equivalence between the two approaches has been
demonstrated; see \cite{ZBS14}.
They estimated the total magnetic helicity density of the active region
AR~11158 by multiplying the total magnetic helicity density,
$\int\HM(k)\,\dd k$, with the volume spanned by the surface area
of the magnetogram of $186\times186\Mm^2$ and an assumed height of $100\Mm$.
In this way, they found a total magnetic helicity of $10^{43}\Mx^2$,
which agrees with the value found by several groups
\citep{VAMC12,LS12,Jing12,Georg13}.
We recall that $1\Mx=1\G\cm^2$ is the unit of magnetic flux.
The linkage of flux tubes is proportional to the product of the two
fluxes of two interlinked flux tubes and thus has the unit $\Mx^2$.

The work done so far has shown that at the solar surface the magnetic
helicity density is negative in the northern hemisphere and peaks at
$k\approx0.06\Mm^{-1}$, which corresponds to a scale of about $100\Mm$;
see \cite{BPS17}.
Surprisingly, in their work there was no evidence for the sign reversal
that was expected based on theoretical models \citep{BB03} and as was
also seen in the active region AR~11515, which was exceptionally helical
\citep{LYG12,Wang14,ZBS16}.
A positive sign of magnetic helicity has also been seen in the mean-field
computations of \cite{PP14}.
The work of \cite{BPS17} was preliminary in the sense that one
should really perform an analogous analysis using spherical harmonics,
but this has not yet been done and the two-scale formalism has not yet
been developed for that case.
Also, they only analyzed three Carrington rotations of the Sun.
Meanwhile, by analyzing a much larger sample, \cite{Singh18} found many
other Carrington rotations for which the spectrum is bihelical.
However, the energy contained in the large-scale contribution with
opposite sign of magnetic helicity is rather weak.

\subsection{Magnetic helicity in the solar wind}

\begin{figure}\begin{center}
\includegraphics[width=\columnwidth]{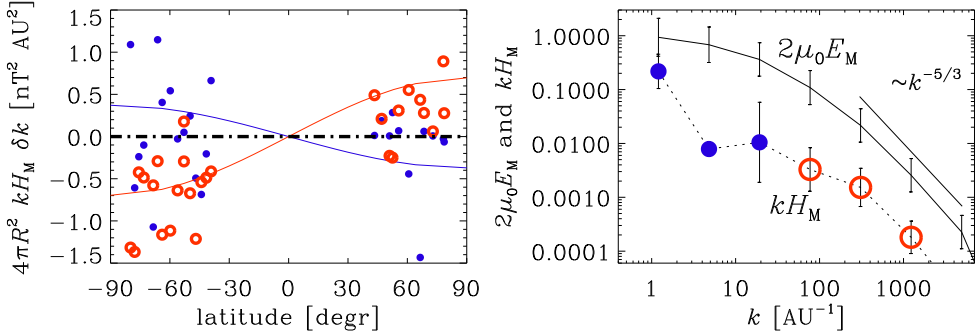}
\end{center}\caption[]{
{\em Left}: Latitudinal dependence of spectral magnetic helicity for
$k=300\AU^{-1}\approx2\times10^{-3}\Mm^{-1}$ (open red symbols) and
$k=1.2\AU^{-1}\approx10^{-5}\Mm^{-1}$ (filled blue symbols).
{\em Right}: magnetic helicity spectrum for heliocentric distances above
$2.8\AU$ for the northern hemisphere, where filled blue symbols denote
negative values and open red ones positive values.
}\label{plat4b_JPP}\end{figure}

To compute magnetic helicity from time series of the three components
of the magnetic field vector in the solar wind, $\BB(t)$, one first
adopts the Taylor hypothesis, i.e., $\BB(r)=\BB(r_0-u_rt)$, where $r$
is the radial coordinate and $u_r\approx800\km\s^{-1}$ is the solar wind
speed in the $r$ direction at high solar latitudes.
Next, one makes use of the isotropic representation of the
Fourier-transformed two-point correlation tensor \citep{Mof78,MG82}
\EQ
\bra{\hat{B}_i(\kk)\hat{B}_j^\ast(\kk')}=\left[
\left(\delta_{ij}-\hat{k}_i\hat{k}_j\right)
2\mu_0\EM(k)-\ii k_l\epsilon_{ijl}\HM(k)
\right]\,{\delta^3(\kk-\kk')\over 8\pi k^2}
\EN
where $\EK(k)$ and $\HM(k)$ are again the magnetic energy and magnetic
helicity spectra.
\cite{MG82} analyzed {\em Voyager} data, but {\em Voyager~1} and {\em 2} were close to the
ecliptic in the data analyzed, so the helicity fluctuated around zero.
The work of \cite{BSBG11} used data from {\em Ulysses}, which flew over
the poles of the Sun.
They showed that $\HM(k)$ changes sign at the ecliptic, as expected,
but it is positive at small scales; see \Fig{plat4b_JPP}.
Thus, we see that the sign of magnetic helicity is the other way around
than what is expected in the dynamo interior and what is found at the
solar surface.

Simple numerical models of \cite{WBM11,WBM12} and \cite{BAJ17} confirm the
sign change of magnetic helicity between the dynamo interior and the halo.
Thus, for the Sun, we expect a similar sign change to occur somewhere
above the surface, and perhaps already within the corona.
Realistic corona simulations by \cite{BBP13} have now shown that
this magnetic helicity reversal occurs when the magnetic plasma beta
drops below unity \citep{BSB18}, i.e., when the plasma becomes dominated
by magnetic pressure compared with the gas pressure.
\cite{BSBG11} explained this reversal by a subdominance of the $\alpha$
effect compared with turbulent diffusion.
An alternative explanation was offered by \cite{WBM12}, who argued that
a turbulent-diffusive magnetic helicity flux down the gradient of the
local magnetic helicity density can result in its sign change, because,
unlike temperature, magnetic helicity density is not sign-definite.
Whether any of these explanations is right needs to be seen through
future work.

The question whether and where the anticipated sign reversal of
magnetic helicity above the solar surface happens can hopefully be
addressed soon using observational techniques.
Several techniques can be envisaged.
There is first of all the {\em in situ} technique by which one puts a
magnetometer into space to determine magnetic helicity, as done for the
data from {\em Voyager} \citep{MG82} and {\em Ulysses} \citep{BSBG11}.
NASA's {\em Parker Solar Probe} mission will have a magnetometer on board
as well and will be able to approach the Sun to within $0.04\AU=6000\Mm$.
If, however, the sign reversal occurs near the point where the plasma
beta is unity, as now predicted by \cite{BSB18}, we would need to measure
even closer to the surface.
This requires remote sensing via polarimetry.
A helical magnetic field corresponds to a rotation of the perpendicular
magnetic field vector about the line of sight.
Therefore, at sufficiently long wavelengths, Faraday rotation
could either enhance or diminish the net Faraday depolarization
that results from the superposition of polarization vectors from
oppositely oriented fields \citep{BS14}.
The application to the Sun was recently explored by \cite{BAJ17}.
To determine magnetic helicity, one needs measurements over a range
of different wavelengths.
Both ESA's {\em Solar Orbiter} mission as well as ground-based
observations with the Daniel K. Inouye Solar Telescope could be
capable of this task using infrared wavelengths.
At longer wavelengths, the Atacama Large Millimeter Array could
be utilized instead.
Again, more detailed estimates are given in \cite{BAJ17}.

\subsection{The solar dynamo dilemma}
\label{DynamoDilemma}

The solar dynamo dilemma was posed by \cite{Par87} in response to the
then emerging helioseismological result that the Sun's internal angular
velocity, $\Omega(r,\theta)$, increases in the outward direction,
i.e., $\partial\Omega/\partial r>0$, where $r$ is radius and $\theta$
is colatitude.
This was found to be the case in the bulk of the convection zone and
especially in the lower overshoot layer, also known as the tachocline.
The Parker--Yoshimura rule for the migration direction of $\alpha\Omega$
dynamo waves states that waves migrate in the direction
\EQ
\xxi_{\rm migration}=-\alpha\pp\times\nab\Omega,
\label{Migration}
\EN
where $\pp$ is the unit vector in the azimuthal direction.
It was based on the original paper of \cite{Par55} and generalized in
a coordinate-independent way by \cite{Yos75}.
Indeed, already the first global and fully selfconsistent convective
dynamo simulations of \cite{Gil83} and \cite{Gla85} showed poleward
migration and this has been confirmed in subsequent simulations;
see, e.g., \cite{KKBMT10}.
Not surprisingly, corresponding mean-field dynamos with selfconsistently
generated differential rotation driven by the $\Lambda$ effect
\citep{Rue80,Rue89,RH04} with magnetically modulated convective energy
fluxes \citep{BMT92} also confirmed this somewhat disappointing result.

Several possible solutions out of the solar dilemma have been proposed;
see the reviews by \cite{SIS06}, \cite{MT09}, and \cite{Cha10}.
\cite{CSD95} have shown that the Sun's meridional circulation can turn
the dynamo wave around and produce equatorward migration owing to the
local circulation speed at the bottom of the convection zone where it
is believed to point equatorward.
This type of model is now referred to as Babcock--Leighton flux transport
dynamo \citep{DC99}, but it can only work if the turbulent magnetic
diffusivity $\etat$ is low enough.
This is already a problem, because $\etat$ should be more than ten times
smaller than what is expected from mixing length theory \citep{Kri84}.
Furthermore, the induction zones of $\alpha$ effect and differential
rotation must be non-overlapping.
This is also not really borne out by simulations.
Indeed, when the induction zones are non-overlapping, meridional circulation
was always found to lead to a suppression of the dynamo,
i.e., the dynamo becomes harder to excite \citep{Rae86a,Rae95}.
Another approach is to adopt a dynamo that attains its
equatorward migration from the near-surface shear layer.
This is a layer in the top $40\Mm$ of the Sun, where
$\partial\Omega/\partial r<0$, which causes equatorward migrating
dynamo waves when $\alpha$ is positive in the northern hemisphere
\citep{Bra05}.
Such a dynamo model has been developed by \cite{PK11} and \cite{Pip17}.

\cite{CS17} have presented an updated version of the one-dimensional
phenomenological dynamo model of \cite{Lei69} by including a number
of effects such as the evolution of the radially integrated toroidal
magnetic field, the latitudinal variation of the surface angular velocity,
turbulent downward pumping, and several other features.
Using surface magnetic field observations, \cite{CS15} showed that the
emerged magnetic flux at the solar surface controls the net toroidal
magnetic flux generated in each hemisphere.
This allowed \cite{CDSS18} to compute maps of poloidal and toroidal
magnetic fields of the global solar dynamo.

Global simulations continue to have a hard time reproducing not only
the near-surface shear layer with $\partial\Omega/\partial r<0$, but
also the approximately spoke-like angular velocity contours
throughout the deeper parts of the convection zone and of course the
equatorward migration of the sunspot belts.
Whether or not they are explicable in terms of the Parker--Yoshimura
rule needs to be seen.

Some of the butterfly diagrams derived from the simulations of
\cite{KMB12,KMCWB13} look convincing, but here an equatorward dynamo
wave results from a local minimum of the differential rotation at
mid-latitudes \citep{WKKB14}.
Another possibility was proposed by \cite{ABMT15}, who also found
equatorward migration.
They argued this to be the result of nonlinearity.
More detailed analysis would be needed to clarify the true reason behind
equatorward migration in the models.
Furthermore. the angular velocities of all these models exceed that of
the Sun by at least a factor of three \citep{BMBBT11}, although simulations
with the EULAG code \citep{GCS10,Racine} seem to produce cyclic solutions
already at the solar angular velocity.
Larger angular velocities were also used by \cite{KMCWB13} and
\cite{KKOWB17}, who compared differences in the parameters used in the
models of different groups.

All the global simulations have certain shortcomings that we need to be
aware of when assessing their overall validity.
Most of the simulations do not yet show well-developed shear
layers, although higher resolution computations, enabling higher density
stratification overall, and especially in the surface regions, have shown
their emergence, even though yet with quite a different appearance as
the observed one \citep[see, e.g.,][]{Hotta14,Hotta15,Hotta16}.
Furthermore, the contours of constant angular velocity are still
distinctly cylindrical and not spoke-like, as found from helioseismology
\citep{Schou}.
Whether this mismatch in the angular velocity contours between simulations
and observations implies also a problem for the solar dynamo remains an
open question, however.
Not only the contours of angular velocity are distinctly cylindrical in
simulations, but also the streamlines of meridional circulation do not
correspond to a single or double cell, as seen in some helioseismic
inversions \citep{Zha13}.
This might not be a problem for the dynamo that is shaped by the
near-surface shear layer, but it would be a problem for the flux
transport dynamo models.

\begin{figure}\begin{center}
\includegraphics[width=.8\columnwidth,angle=-90]{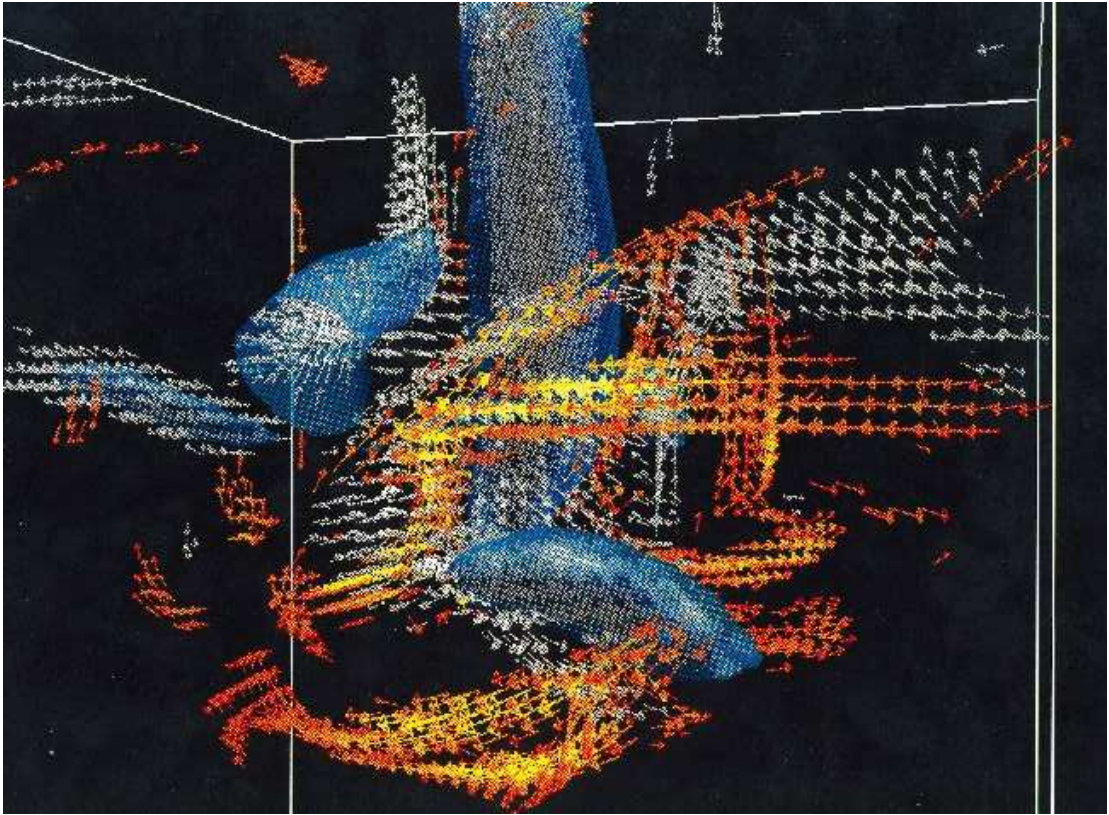}
\end{center}\caption[]{
Vorticity vectors $\oo$ (from gray to white as $|\oo|$ increases) and
magnetic field vectors $\BB$ (from red to yellow as $|\BB|$ increases).
Only vectors whose strength exceeds a threshold of three times the rms
value are plotted.
An isosurface of constant pressure fluctuation is shown in blue and it
is seen to encompass some of the vortex tubes, especially the one around
the cyclonic downdraft descending from the middle of the domain.
Magnetic flux tubes are seen to be wrapped around the spinning downdraft
and are being pushed down, which reflects the effect of downward pumping.
}\label{swirling_Btubes}\end{figure}

\subsection{Flux transport dynamos}

A popular scenario for the solar dynamo is the flux transport dynamo.
It emerged as a remarkable finding when \cite{CSD95} extended earlier
studies of \cite{Rae86a} regarding the effects of meridional circulation
on the dynamo.
In the original work of \cite{Rae86a}, the induction zones corresponding
to $\alpha$ effect and differential rotation were overlapping, and he
found that meridional circulation always has a suppressing effect on
the dynamo, which eventually became non-oscillatory.
However, when the two inductions zones were separated such that the
$\alpha$ effect operates only near the surface and differential rotation
only at the bottom of the convection zone, the solutions remained
oscillatory and a new dynamo mode appeared.
It is still oscillatory, with dynamo waves migrating in the direction
of the meridional circulation---regardless of what was predicted by the
Parker--Yoshimura rule; see \Eq{Migration} in \Sec{DynamoDilemma};
see \cite{KRS01} for more thorough studies of the dynamo properties.

Further fine-tuning of this approach has now resulted in
detailed models that can reproduce the equatorward migration
of the solar dynamo, the polar branch, and the cycle period
\citep{DC99,DG01,NC02,DG06,DdTGAW04,DdTG06,CNC04,CC06,NMM11}.
It requires, however, low turbulent diffusivities of about ten times
below what is estimated based on mixing length theory.
It also requires the existence of tilted flux tubes rising to the surface
to motivate the occurrence of an $\alpha$ effect at the surface only.
By contrast, in conventional models, $\alpha$ peaks in the lower part
of the convection zone; see figure~2b of \cite{BT88}.
Also, the pattern of meridional circulation should ideally be a single
cell, although multiple cells could also be possible as long as the flow
in the tachocline is equatorward \citep{HKC14}.

The idea of a flux transport dynamo is hard to reconcile with dynamo
theory and global simulations, which predict distributed induction zones,
larger convections speeds and therefore larger turbulent diffusivities,
and a time-dependent meridional circulation pattern that is aligned with
the rotation axis.
The latter feature is not observed in the Sun---casting therefore some
doubt on the predictions from simulations.
Smaller diffusivities could be explained by smaller-scale convection cells.
We return to this in \Sec{Deardorff} on the convective conundrum.
The idea of an $\alpha$ effect operating only near the surface could
perhaps be reconciled with theory if $\alpha$ was vanishingly small in
the interior and only nonvanishing near the interface to the outer corona.
But these are just speculations that have no theoretical basis.
Therefore, the flux transport dynamo appears to be in many ways the
result of some intelligent design, without footing in the theory of
hydromagnetic turbulence.

\subsection{Downward pumping versus turbulent diamagnetism}
\label{DownwardPumping}
\label{DoesMagnetic}

Downward pumping was clearly seen in the numerical dynamo simulations of
turbulent convection; see \Fig{swirling_Btubes}, which is similar to those
of an early review on this by \cite{BT91} and the work of \cite{NBJRRST92}
and \cite{BJNRST96}.
\cite{TBCTK98} and \cite{TBCTK01} quantified many aspects of pumping
in dedicated numerical experiments.

In the simulations mentioned above, the dynamical range
of the correlation time $\tau=(\urms\kf)^{-1}$ is not yet sufficiently
large, so $\tau$ does not change significantly between top and bottom
of the domain.
Therefore, the difference between
\EQ
\gamma=\left\{
\begin{array}{ll}
-{1\over6}\tau\nab\overline{\uu^2} &
\mbox{(if $\tau$ is outside the gradient),}\\
-{1\over2}\nab({1\over3}\tau\overline{\uu^2})\equiv-\half\nab\etatz &
\mbox{(if $\tau$ is under the gradient),}\\
\end{array}
\right.
\EN
is not yet significant.
Theoretically, it is not clear which of the two formulations is the
correct one.
The former version was obtained by \cite{Rae69c}, but a variation of
$\tau$ was not explicitly considered.
The latter version was obtained by \cite{RS75}.
Near the surface of the Sun, $\etatz$ increases with depth \citep{Kri84},
so $\gamma=-\half\nab\etat$ would point upward, but $\overline{\uu^2}$
decreases with depth, so $\gamma=-{1\over6}\tau\nab\overline{\uu^2}$
would point downward, which would be in agreement with the simulations.

This question has implications on whether or not the $\gamma$ effect can
be understood as turbulent diamagnetism, because we could then write
\EQ
-\ggamma\times\meanBB-\etat\mu_0\meanJJ=
-\etat^{1/2}\,\nab\times\left(\etat^{1/2}\,\meanBB\right),
\EN
where $\etat^{1/2}$ would play the role of both a renormalized magnetic
diffusivity and a renormalized magnetic permeability.

Mean-field simulations have long shown a significant effect of
pumping on the migration of the dynamo wave \citep{Kit91,BMT92b}.
Significant equatorward pumping near the surface and poleward pumping
deeper down was recently found in global test-field calculations
\citep{War18}.
This seems to be contrary to what was assumed in some flux-transport
dynamos and would be more advantageous for models where the equatorward
migration of the dynamo wave resulted from flow conditions nearer to
the surface.

There is also topological pumping \citep{DY74}.
It has been applied to convection, where the up- and downflows tend
to occupy distinct regions in each horizontal plane.
The effective pumping velocity depends only on the vertical flow in horizontally
connected regions, which we refer to as flow lanes.
For example near the surface we have horizontally connected downflow
lanes, so pumping would be downward.
In the deeper layers, however, the downdrafts are isolated and the
upwellings are horizontally connected, so topological pumping would here
be upward.
Numerical simulations have confirmed this effect \citep{Art83}
and have been applied to what is known as the fountain flow in galaxies
\citep{BMS95}.

As seen above, many of the turbulent transport coefficients have both
kinetic and magnetic contributions.
For example, the $\alpha$ effect has both kinetic and current helicities,
and the turbulent pumping effect also has two contributions, namely
\EQ
\gamma=-{1\over6}\tau\nab(\overline{\uu^2}-\overline{\bb^2}/\mu_0\rho_0),
\EN
but the turbulent magnetic diffusivity has only one, i.e.,
$\etat=\onethird\tau\overline{\uu^2}$.
This was been shown by \cite{RKR03}; see also \cite{BS05} for a review.
However, one should be aware that this result is a consequence
of the second order correlation approximation and the assumption of
isotropy, and has not yet been confirmed with the test-field method.
It is simply another one of the many open question in mean-field theory.

\subsection{Contributions to the $\alpha$ effect}

There is a related uncertainty regarding the $\alpha$ effect.
In the original derivation of \cite{SKR66}, $\alpha$ was proportional
to the gradient of $\ln\rho\urms$.
The $\alpha$ effect also depends on the angular velocity, so the full
expression can then be written in the form
\EQ
\alpha=-\ell^2\OO\cdot\nab\ln(\rho^\sigma\urms),
\EN
where $\ell$ is the correlation length of the turbulence and
$\sigma$ is an exponent that characterizes the importance of
density stratification relative to velocity stratification.
\cite{RK93} confirmed $\sigma=1$ for rapid rotation,
but found $\sigma=3/2$ for slow rotation.
Recent work using the test-field method now shows that $\sigma=1/2$ for
forced turbulence and convection with strong density stratification, while
for supernova-driven turbulence $\sigma=1/3$ was found \cite{BGKKMR13}.
In any case, contrary to the earlier scaling, $\sigma$ is always less
than unity.

We clearly see that at the equator, the rotation and stratification
vectors are at right angles to each other, so $\alpha=0$.
It is important to realize, however, that a nonvanishing $\alpha$
is in principle also possible at the equator if $\alpha$ is
the result of an instability, whose eigenfunctions are helical.
The signs of helicity and $\alpha$ effect depend then on the initial
conditions.
This has been demonstrated both for the magneto-buoyancy instability
\citep{CMBR11} and for the Tayler instability \citep{Gellert,BBDSM12}.
Even though the growth rates are the same for both signs of helicity,
only one sign will survive in the nonlinear regime owing to what is
called mutual antagonism in the related application of spontaneous
chiral symmetry breaking leading to finite handedness of biomolecules
\citep{Frank}.
This is believed to be relevant to time at the origin of life on Earth
\citep{San03,BM04,BAHN}.

The presence of $\alpha$ in a system affects also the turbulent magnetic
diffusivity.
This was not theoretically expected, but it is easy to see that such a
term is theoretically possible.
\cite{BSR17} showed that, for intermediate values of $\Rm$, $\etat$
{\em decreases} by almost a factor of two.
This may not be very much in view of other uncertainties known in
astrophysical turbulence, but it can be important enough to make a
difference in theoretical studies, where reasonably accurate estimates
of turbulent diffusivity are now available.

\subsection{Buoyant flux tubes}

The notion of flux tubes was quite popular since Parker's other early
work of 1955, when he argued that bipolar regions at the solar surface
can be explained by flux tubes piercing the surface.
This appeared quite plausible, given that the anticipated depth of those
flux tubes was expected to be about $20\Mm$ \citep{Par55b}.
In that case, the depth of flux tubes and the separation of bipolar
regions would be comparable, but in subsequent years, \cite{Par75} argued
for a storage depth of magnetic flux tubes of about $200\Mm$,
which is the bottom of the convection zone.
This makes the flux tube picture much harder to accept, because flux
tubes not only expand during their ascent, but their dynamics is
rather complicated and by no means as simple as that of a garden hose
sweeping through the air and then piercing the roof of a tent.
This was demonstrated in numerous simulations
\citep{Fan01,Fan08,Fan09,HAGM09,SAGT15}.

\begin{figure}\begin{center}
\includegraphics[width=.9\columnwidth,angle=0]{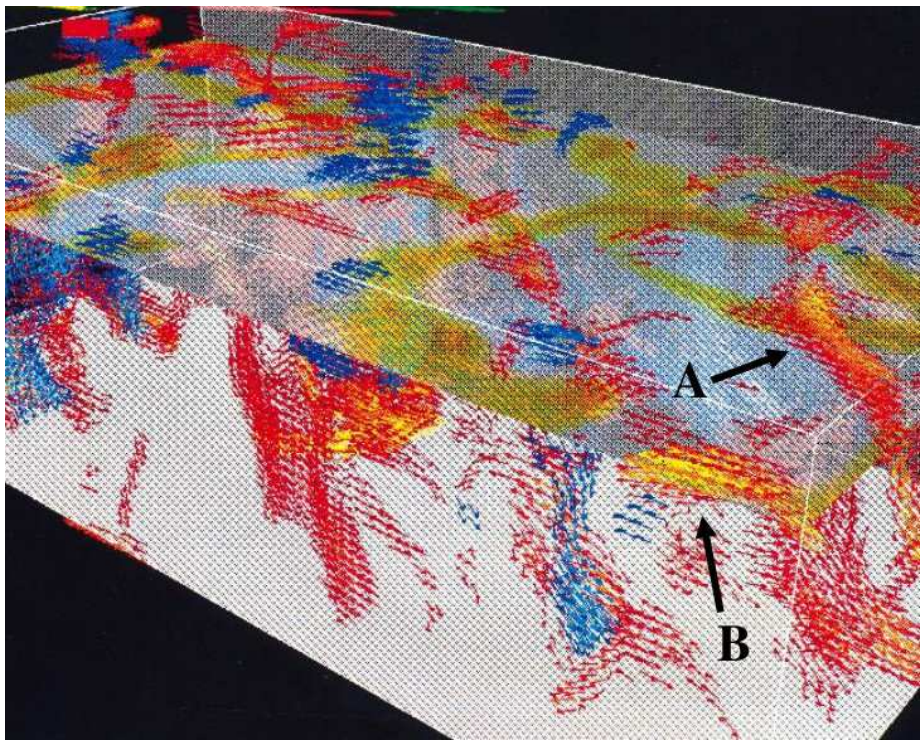}
\end{center}\caption[]{
Magnetic field vectors (red to yellow) and vorticity vectors (blue) for
a convectively driven dynamo in an elongated domain with a radiative
cooling layer above a surface marked with a transparent visualization
of temperature.
Note the appearance of flux tubes crossing the surface (see the positions
marked with A and B).
}\label{rising_tube}\end{figure}

Some successes of the flux tube picture have however been noted.
In some of those cases, the magnetic flux tubes are analogous to the
vortex tubes seen in the direct numerical simulations of \cite{SJO90}.
The meshpoint resolution of $96^3$ used at the time was moderate
by nowadays standards.
In \Fig{rising_tube} we reproduce a snapshot from a dynamo simulation
similar to those of \cite{BJNRST96}, where a cooling layer was included
at the top (in addition to an overshoot layer at the bottom of the
convectively unstable layer).
One sees buoyant flux tubes having reached the surface in various places.
However, saying that these are the tubes that make a sunspot pair would
be rather optimistic, because those magnetic tubes are analogues to the
vortex tubes in turbulence and have radii comparable to the resistive
length \citep{BPS95}, so they only look solar-like because those
simulations did not yet have large resolution.

In the visualizations discussed above, flux tubes were identified
as coherent assemblies of mutually aligned vectors whose strengths
exceeds a certain threshold of typically three times the rms value of
the magnetic field.
This has the advantage that those flux structures are dynamically
important and would affect the gas pressure balance to produce
magnetic buoyancy, as was demonstrated in figure~10 of \cite{BJNRST96}.
Obviously, those flux tubes terminate when the field becomes weak,
even though the magnetic field lines continue.
By visualizing field lines integrated along any local field
vectors---regardless of their strength, \cite{NBBMT13,NBBMT14}
and \cite{NM14} were able to demonstrate the existence of serpentine
structures encompassing much of the solar circumference.
In weak sections of the structure, its dynamics is governed by advection
rather than magnetic buoyancy.
Rising structures automatically expand while descending ones contract,
so most of the magnetic buoyancy was found to operate in descending
structures; see, again, figure~10 of \cite{BJNRST96}.
It is therefore difficult to judge whether visualizations of integrated
field lines can tell us much about Parker's original picture of producing
bipolar regions in the Sun.

One more point is in order here.
The idea about flux tube storage mentioned by \cite{Par75} is an aspect
that has not been verified nor is it seen in simulations; see those
of \cite{GK11} for an attempt to amplify magnetic flux at the bottom of
the convection zone.
An important ingredient of flux transport dynamos is the induction effect
at the surface that is supposedly caused by the decay of tilted active
regions \citep{Bab61,Lei69}.
If these effects really operate, one should be able to verify them in
a dedicated simulation using the test-field method.
This has not yet been done.

\subsection{Surface flux transport models}

In spite of the problems encountered in modeling the solar dynamo,
there has been some success in modeling the advection of active regions
using what is called the surface flux transport model \citep[see,
e.g.,][]{HGHA15}.
This is a two-dimensional model that ignores the dynamics in the vertical
direction.
That this actually works is remarkable and suggests that active
regions just ``float'' at the surface.
Such models are perhaps the best we have to predict the magnetic field
after it disappeared on the far side of the Sun.
Of course, it is not a model of the solar
dynamo because it assimilates continuous input from observations.

The fact that active regions appear to {\em float} at the solar surface
might well be consistent with them being locally maintained entities at
or just beneath the surface.
The one process that is known to lead to magnetic flux concentrations
of that type is the negative effective magnetic pressure instability
(NEMPI); see \cite{BRK16} for a review.
This is a mean-field process in the momentum equations, where
the Reynolds and Maxwell stresses attain a component proportional
to $\meanBB^2/2\mu_0$, which acts effectively like a negative
pressure by suppressing the turbulent pressure; see \cite{vanBall84}
for early ideas along similar lines of thought.
Mean-field investigations started with \cite{KRR89,KRR90,KMR93,KRR95,KMR96}
and \cite{KR94}, while the first simulations of the mean-field equations
were produced by \cite{BKR10,BKKR12} and \cite{KBKR12}.
This effect was also detected in various direct numerical simulations
\citep{BKKMR11,KeBKMR12,KBKMR13}.
The formation of bipolar regions from NEMPI was first studied by
\cite{WLBKR13,WLBKR16}, who allowed for the effects of an overlying
corona.

NEMPI has a number of properties that negatively affect its role
in explaining magnetic flux concentrations in the Sun.
One is rotation: already rather small Coriolis numbers well below unity
suppress the instability \citep{Los12,Los13}.
NEMPI was found to be not excited in convection \citep{KBKKR16}, which
was possibly due to insufficient scale separation in their simulations.
However, more detailed work showed that the derivative of the effective
magnetic pressure with respect to the mean magnetic field was found to
have an unfavorable sign for the onset of NEMPI.
Furthermore, radiation transport was found to make the onset of NEMPI
oscillatory and the horizontal length scale of the eigenfunctions about
ten times smaller \citep{PB18}.
Even if NEMPI were excited, the flux concentrations would be too weak
to produce sunspots.

An alternative possibility that has been discussed in the past is
the suppression of the convective heat flux by magnetic fields.
This could lead to a large-scale instability \citep{KM00}.
Unfortunately, not enough is known about this possibility,
nor has it been detected in direct numerical simulations as yet.
Eigenvalue calculations of M.\ Rheinhardt (unpublished) suggest that
this crucially depends on the nature of the radiative boundary condition
imposed at the top.
This clearly needs to be addressed further to find out whether this
instability is a real phenomenon or perhaps even an artifact of this
boundary condition.

\subsection{The convective conundrum}
\label{Deardorff}

Over the past few decades, numerous simulations have
demonstrated how difficult it is to reproduce the Sun
\citep{Gil83,Bru04,BMT04,BMBBT11,NBBMT13}; see also \cite{MT09} for a review.
If it is true that the solar dynamo is driven by the velocity field
in the Sun, one wonders what exactly is ``wrong'' with it.
That something is not quite right is immediately evident when
comparing the contours of constant angular velocity from helioseismology
with those from simulations; see \cite{TCDMT03} for a review and the
discussion in \Sec{DynamoDilemma}.
We return to current proposals of resolving this problem further below.

A more subtle discrepancy is that the horizontal scales of convection
are observed to be much smaller than what is seen in convection.
This phenomenon came to be called the convective conundrum \citep{OMFA16}.
Global convection simulations of \cite{Miesch08} predict giant cells
that are not observed.
Helioseismological observations with the time-distance method
predict very low velocities at those scales \citep{HBBG10,HDS12,HGS16},
but this, in turn, could also be an artifact of excessive noise reduction.
This was argued by \cite{Greer}, who finds significantly larger velocities
at the theoretically expected levels using the ring diagram local
helioseismology technique.

From a theoretical point of view,
one problem is that all global simulations of convection assume a
prescribed unstable layer of about $200\Mm$ depth.
This may not be realistic, because of the effects of intense downdrafts
driven by surface cooling \citep{Spr97}.
He found that the deeper layers would remain always
convectively unstable, but subsequent work suggested that the deeper
layers are convecting only because of strong mixing driven by the surface
motions \citep{Bra16,Kapy_etal17}.
Thus, the depth of the convection zone should be a sensitive function
of the vigor of convection in the surface layers.

The deeper layers may not transport the convective flux based on the
local superadiabatic gradient, as assumed in standard mixing length
theory \citep{Vit53}, but based on another term suggested first by
\cite{Dea66,Dea72} in the geophysical context and applied to the solar
context by \cite{Bra16}.
The calculation is analogous to that presented in \Sec{TauApprox},
but instead of \Eqs{inductb_approx}{dudt_approx}, we now have
\EQ
{\partial s\over\partial t}=-\uu\cdot\nab\meanS+...\;,\quad\mbox{and}\quad
{\partial\uu\over\partial t}=-\grav s/\cP+...\;,
\label{dsdt_approx}
\EN
where $S=\meanS+s$ is the specific entropy separated into mean and
fluctuating parts, $\grav$ is gravity, and $\cP$ is the mean specific
heat at constant pressure.
Computing the correlation $\meanFFFF=\overline{s\uu}$, which is
proportional to the mean convective energy flux, we have, analogously
to \Eq{TwoTerms}, two terms that are here
\EQ
{\partial\meanFFFF\over\partial t}
=\overline{\uu\dot{s}}+\overline{\dot{\uu}s}.
\EN
The first ones leads to the usual negative gradient contribution,
$-\tau\,\overline{u_iu_j}\,\nabla_j\meanS$, but there is a second term,
$-\tau\grav\overline{s^2}/\cP$, which is the Deardorff term; see
\cite{Bra16} for details.
This term is always in the negative direction of gravity and
 proportional to the square of the specific entropy fluctuation.
The enthalpy flux is thus the sum of a gradient term proportional to
the usual superadiabatic gradient and a Deardorff term.

A full mean-field model of the Sun must include hydrodynamics and
thermodynamics \citep{BMT92,Rem05}.
Such models were considered by \cite{TR89}, who found what appeared to be
a new instability of the full system of equations; see \cite{RS92} for its
detailed investigation.
However, this turned out to be essentially a Rayleigh-Ben\'ard type
instability \citep{TBMR94}.
It could potentially be stabilized by having a turbulent viscosity and
a turbulent thermal diffusivity that are large enough.
Alternatively, of course, it could be stabilized by a sufficiently small
or even negative superadiabatic gradient, which would naturally occur
in Deardorff-type convection discussed above.

Global simulations using a more realistic opacity prescription
result in extended subadiabatic layers \citep{KVKB18}.
They also lead to significant latitudinal specific entropy gradients,
which are known to alleviate the tendency to form cylindrical contours
of constant angular velocity arising from the Taylor-Proudman theorem
\citep{Rue89,BMT92}.
Clearly, more work in that direction is needed to clarify the role and
origin of these extended subadiabatic layers.

\subsection{Solar equatorward migration from an oscillatory $\alpha^2$ dynamo}
\label{SolarEquatorward}

Another idea that has been discussed is that the equatorward migration
could be caused by an $\alpha^2$ dynamo.
\cite{SG03} found oscillatory $\alpha^2$ dynamos for a nonuniform
$\alpha$ distribution in the radial direction.
Later, \cite{MTKB10} found an oscillatory $\alpha^2$ dynamo with
equatorward migration in a model with a change of sign of $\alpha$
across the equator.
It was therefore thought that a gradient in the kinetic helicity was the
reason behind the oscillatory nature of the dynamo and thus equatorward
migration.
\cite{KMCWB13} investigated the phase relation between toroidal and poloidal
magnetic fields in their oscillatory convectively driven dynamo with
equatorward migration and found a phase shift of $\pi/2$, which is
compatible with what is expected for an oscillatory $\alpha^2$ dynamo.
\cite{MS14} confirmed this finding for a dynamo in Cartesian geometry
and reinforced the suggestion that the solar dynamo might indeed be of
$\alpha^2$ type.
Then, \cite{CBKK16} found that the oscillatory $\alpha^2$ dynamo requires
highly conducting plasma at high latitudes or, alternatively, a perfectly
conducting boundary condition at high latitudes, as is often assumed in
spherical wedge simulations \citep{MTBM09}.
This was then confirmed through the realization that an oscillatory migratory
$\alpha^2$ dynamo is possible even with constant $\alpha$ effect provided
there are two different boundary conditions on the two sides \citep{Bra17}.
With this realization, the idea of a solar $\alpha^2$ dynamo now begins to
sound somewhat artificial.
The best use of such a model might therefore now be the
application to the study of magnetic helicity fluxes, as discussed in
\Sec{MagneticHelicityFluxes}.

\section{Stellar dynamos}

Cycles like the 11 year sunspot cycle are known to exist on other main
sequence stars with outer convection zones.
Stellar activity cycles are usually detected in the calcium H and K lines
which form in chromospheric magnetic loops in emission \citep{Wil78}.
This was already known since the early work of \cite{ES13}.
Some cycles are also seen in X-rays and in extreme ultraviolet,
for example that of $\alpha$ Cen A \citep{Ayr09,Ayr15}.
For some stars, it has also been possible to observe surface magnetic
fields directly through Zeeman Doppler imaging.
An example is HD~78366, where it has been possible to see a sign reversal
of the magnetic field on a $\sim2$ years timescale \citep{Morgenthaler11},
which was not evident from just the times series \citep{BMM17}.
Unfortunately, Zeeman Doppler imaging requires many nights on big
telescopes with high-resolution spectrographs.
It then becomes prohibitive to cover many epochs, which is a serious
disadvantage over the more regularly spaced light curve observations.
On the other hand, neither circular nor linear polarization has been
detected on $\alpha$ Cen A, indicating the absence of a net longitudinal
magnetic field stronger than $0.2\G$ \citep{Koc11}, which remains puzzling.

\subsection{Stellar cycle frequency, rotation, and activity}

It has been known for some time that stellar activity increases with
increasing rotation rate up to a certain point above which the activity
saturates.
However, to be able to compare different stellar types with different
convective turnover times ranging from $\tau=7$ to 26 days between F7
and K7 dwarfs, it was found to be useful to normalize the rotation period
by $\tau$.
Indeed, the dependence of stellar activity on the rotation period
$P_{\rm rot}$ is well described by $P_{\rm rot}/\tau$ \citep{Vil84,Noyes84},
which is referred to as the Rossby number in stellar astrophysics.
Note, however, that in astrophysical fluid dynamics the inverse Rossby
number or Coriolis number is defined as $2\Omega\tau$, which is larger
than $\tau/P_{\rm rot}$ by a factor of $4\pi$ because of
$P_{\rm rot}=2\pi/\Omega$ and the factor of two in the Coriolis force.

Another source of discrepancy is connected with the definition of $\tau$.
In observational stellar astrophysics, one routinely uses the turnover
time at a depth of approximately one pressure scale height above the
bottom of the convection zone.
This works well in the sense that the Rossby number defined in that way
is found to control the chromospheric stellar activity with relatively
little scatter \citep{Noyes84}.
In global simulations, one often uses the rms velocity based on the
entire convection zone together with a rudimentary estimate of the
wavenumber of the energy-carrying eddies; see \cite{KMCWB13}.
Thus, because of these differences, it may well be possible that
theoretical and observational Rossby numbers need to be calibrated
relative to each other.

Indeed, it is unclear how large the
Rossby number of the Sun really is, because solar-like differential
rotation is currently only obtained for somewhat faster rotation rates
than what is expected based on the actual numbers.
According to observations, the transition point may be at
$P_{\rm rot}/\tau\approx2$, but simulations suggest that this happens
at about the angular velocity of the Sun.

Let us now turn to the cycle frequency.
Early work of \cite{NWV84} indicated that the cycle frequency,
$\omega_{\rm cyc}=2\pi/P_{\rm cyc}$, with $P_{\rm cyc}$ being the activity
cycle period (not the magnetic Hale cycle period), increases with rotation
frequency $\Omega=2\pi/P_{\rm rot}$ like a power law,
\EQ
\omega_{\rm cyc}\propto(\Omega\tau)^\nu,
\EN
with $\nu=1.25$.
Using simple dynamo models in a one-mode approximation,
they compared three different nonlinearities ($\alpha$ quenching,
quenching of differential rotation, and magnetic buoyancy), and found
that only the magnetic buoyancy nonlinearity was within certain limits
compatible with the observational result.
By contrast, \cite{KRS83} found an almost perfect agreement with a linear
free wave model which maximizes the growth rate.
However, this model remained unsatisfactory, because it is natural that
a dynamo is nonlinearly saturated.

In another approach, \cite{BST98} argued that both $\alpha$ and $\etat$
are nonlinear functions of the modulus of the magnetic field $B$ of the
form $\propto|B|^n$ and $\propto|B|^m$, respectively.
Again, their models were based on a one-mode approximation.
Interestingly, when such a model is solved without this restriction,
it no longer reproduced the same result.
Regarding magnetic buoyancy, it is important to emphasize that the
modeling of this phenomenon in the one-mode approximation is necessarily
{\em ad hoc}.
In the two-dimensional models of \cite{MTB90}, magnetic buoyancy was modelled
as a mean upward drift, i.e., as a $\BB$-dependent $\gamma$ effect.
This was an idea that was communicated to the authors by K.-H. R\"adler.
The consequences for the cycle period are not known however.
\cite{BST98} argued therefore that the one-mode assumption might not
actually be a ``restriction,$\!$'' but a physical feature of such a model.
This can qualitatively be explained by models with spatial nonlocality,
where only the lowest wavenumbers contribute to $\meanEMF$ in Fourier space.

\subsection{Antiquenched stellar dynamos}
\label{AntiquenchedStellarDynamos}

The reason for the anticipated antiquenching is easily understood when
one considers the expression for the cycle frequency of an $\alpha\Omega$
dynamo \citep{Sti74}
\EQ
\omega_{\rm cyc}\approx\sqrt{\alpha\Omega'},
\EN
where $\Omega'=\dd\Omega/\dd r$ is the radial angular velocity gradient.
Assuming furthermore that $\alpha\approx\Omega\ell$ with $\ell=\ell(B)$
being an effective correlation length and $\Omega'=g\Omega/r$ with
$g(B)$ being a nondimensional shear gradient, we see that
$\omega_{\rm cyc}/\Omega=\sqrt{g\ell/r}$ is independent of $\Omega$
and depends only on the magnetic field, providing thereby a direct
representation of $\alpha$ quenching.

\begin{figure}\begin{center}
\includegraphics[width=\columnwidth]{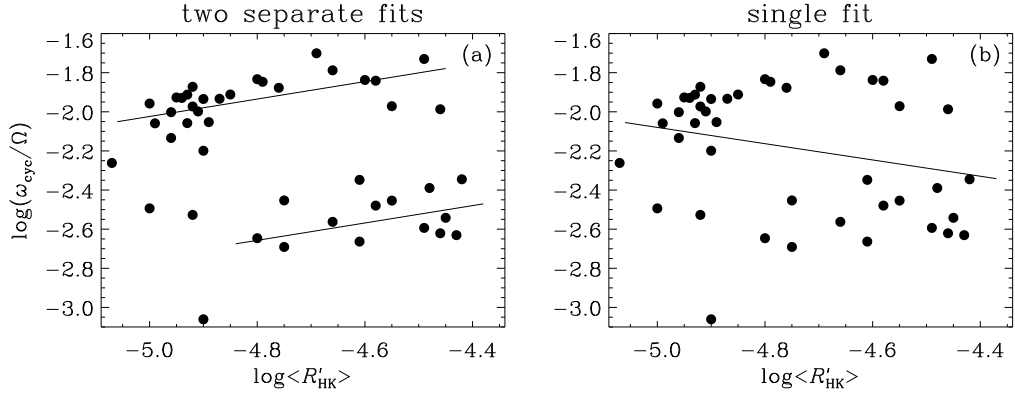}
\end{center}\caption[]{
Cycle to rotation frequency ratios for all primary and secondary cycles
versus $R'_{\rm HK}$
discussed in \cite{BMM17} along with their two separate fits for long and
short cycles (left) compared with the same frequency ratios and a general
single fit through all cycle ratios (right).
}\label{pCycleOutput}\end{figure}

The magnetic activity of late-type stars is usually measured by the
normalized chromospheric Ca~{\sc ii}~H+K line emission, $R'_{\rm HK}$
\citep[e.g.,][]{Vil84,Noyes84}.
Furthermore, the work of \cite{Sch89} has shown that
\EQ
R'_{\rm HK}\propto(B/\Beq)^\kappa
\EN
with $\kappa\approx1/2$; see also \cite{Sch83}.
Therefore, measuring the slope $\nu$ in the representation of
$\omega_{\rm cyc}/\Omega\propto R_{\rm HK}^{\prime\;\mu}$ gives
us insight into the quenching dependence of $\alpha(B)$.
\FFig{pCycleOutput}(a) shows the frequency ratio $\omega_{\rm cyc}/\Omega$
with two separate fits, as proposed by \cite{BST98,BMM17}.
Since $\omega_{\rm cyc}/\Omega$ increases with increasing values of
$R'_{\rm HK}$, i.e., since $\nu>0$, the exponent $n$ must also be positive.
Specifically, we have $n=2\nu\kappa\approx\nu$.
Observations indicate that $\nu\approx0.5$, and therefore also
$n\approx0.5$, but it could be somewhat larger if $g$ increases with
$\Omega$, which is an additional complication that can in principle be
accounted for; see \cite{Bra98b} and \cite{BST98} for details.

The exponent $m$ is constrained by the balance between the destabilizing
contribution, which, for an $\alpha\Omega$ dynamo, is again proportional to 
$\sqrt{\alpha\Omega'}\propto|B|^{n/2}$, and the dissipating contribution
proportional $\etat/L^2\propto\tau^{-1}\propto|B|^m$.
Since $\tau$ enters in the expression for the Rossby number,
$P_{\rm rot}/\tau$, which is proportional to $R_{\rm HK}^{\prime\;\mu}$
with $\mu\approx1$ \citep{BST98}, we have $m=(\nu+1/\mu)\,\kappa\approx0.75$.

As is clear from the explanations above, theoretical models reproduce
a growing $\omega_{\rm cyc}/\Omega$ ratio with increasing $|B/\Beq|$
only with antiquenching and nonlocality.
However, this does not happen in the usual mean-field dynamo models,
where neither of the two effects are included.
Also, three-dimensional global convective dynamo simulations
\citep{SBCBdN17,War18} do not reproduce this trend, which
is why they argue that the correct representation has actually
a negative slope in the $\omega_{\rm cyc}/\Omega$ versus
$R_{\rm HK}^{\prime}$ diagram, as shown in \Fig{pCycleOutput}(b).
To resolve this conflict, more accurate cycle data are needed to be
able to tell whether the correct slope in \Fig{pCycleOutput} is positive
or negative.
This uncertainty is caused by the fact that there is no agreement between
observations and simulations when there are two distinct branches with
a positive slope instead of just one with a negative slope.

\cite{BV07} plotted not the $\omega_{\rm cyc}/\Omega$ ratio,
but $2\pi/\omega_{\rm cyc}\equiv P_{\rm cyc}$ versus
$2\pi/\Omega\equiv P_{\rm rot}$ and found an approximately linear slope,
which would suggest that the $\omega_{\rm cyc}/\Omega$ ratio would
actually be constant, i.e., $\nu=0$ instead of $\nu=0.5$, as found from
almost the same data.

She also suggested that the two branches could correspond to two dynamos
operating simultaneously at two different locations.
Evidence for different dynamo modes in a convection simulation was
presented by \cite{Kapy16,BSCC16}.
This interpretation was also adopted by \cite{BMM17}, who found that
many stars with ages younger than $2.3\Gyr$ might exhibit both ``short''
and ``long'' cycles.
Here the meanings of short (1.6--21 years) and long (5.6--21 years) are
relative and depend on the observed $R'_{\rm HK}$ value.
They examined altogether 11 stars with double cycles.
They also computed cycle periods based on the observed $R'_{\rm HK}$ and
$P_{\rm rot}$ values that would be expected if the cycle periods would
fall exactly onto each of the two branches.
In some cases, it became clear that secondary periods could not have
been observed because the cadence was too long or the time series was
not long enough.
The stars on the two branches with larger and shorter cycle periods have
traditionally also been referred to as active and inactive branch stars.
This interpretation can be justified by noting that longer (shorter) cycle
periods are more (less) pronounced when $R'_{\rm HK}$ is larger.

In addition to the two branches discussed above, there is also another
branch for superactive stars, where $\omega_{\rm cyc}/\Omega$ does indeed
decline with increasing activity.
All the convectively driven dynamo simulations in spherical shells seem
to reproduce this branch qualitatively rather well.
Indeed, one could argue that none of those models reflects the Sun and
that it really operates in a different regime than what has been studied
in spherical shell models so far, where one mainly sees a declining trend.
However, looking again at figure~7 of \cite{War18}, there is actually a
short interval between the stars with antisolar-like differential rotation
(his $\log\Co=0.2$) and the declining branch (his $\log\Co=0.7$), where
the data points are compatible with an increasing trend, albeit with
more noise.

A recent reanalysis of the Mt.\ Wilson data by \cite{OLKP18}
now suggests that many of the double cycles may not be real.
This conclusion was also reached recently by \cite{BoroSaikia}.
Furthermore, according to these recent papers,
the active branch collapses to a circular cloud of
points with no significant slope.
The claim of multiple cycles of stars with different cycle periods
on both branches is argued to be spurious.
The method of \cite{OLKP18} represents a marked methodological improvement
of stellar cycle detection and will need to be looked at more seriously.
On the theoretical side, it would be useful to determine synthetic
light curves to see whether double cycles can occur from modes with
nonaxisymmetric magnetic fields expected for more rapid rotation.

\subsection{Antisolar differential rotation}

The fact that the Sun's differential rotation is as it is, namely
``solar-like'' with a fast equator and slow poles is, in hindsight,
somewhat surprising.
Antisolar rotation has occasionally been seen in numerical simulations
\citep{Gil77,RBMD94,DSB06} and has been associated with a dominance
of meridional circulation \citep{KR04}.
In fact, even simulations that are performed at the nominal solar rotation
rate \citep{BMBBT11} have produced antisolar-like differential rotation,
i.e., the equator rotates more slowly than the poles.
Thus, it seems that there is something about the solar models that makes
them being shifted in parameter space relative to the actual position
of the Sun \citep{Mie15}.
On the other hand, although we are able to reproduce solar-like
differential rotation with a three-fold or five-fold larger Coriolis
number \citep{BMBBT11}, there are still other aspects that are not yet well reproduced,
for example the equatorward migration of the sunspot belts or the contours
of constant angular velocity.

Simulations of \cite{KKKBOP15} have shown that the magnetic activity
increases again at low rotation rates, because the differential rotation
becomes antisolar-like and that the absolute value of this differential
rotation exceeds that of stars with solar-like differential rotation.
There are now indications from the stars of the open cluster M67 that
show an increasing trend for decreasing Coriolis numbers, supporting
the qualitative predictions of the spherical global dynamo simulations
\citep{GBCSH17,BG18}.
Unfortunately, no direct evidence for antisolar-like differential
rotation on dwarfs is available as yet.
With longer time series it might become possible to detect antisolar
differential rotation through
changes in the apparent rotation rate that would be associated with spots
at different latitudes; see \cite{RA15} for details.
So far, antisolar DR has only been observed in some K giants
\citep{SKW03,WSW05,Kovari_etal15,Kovari_etal17} and subgiants
\citep{Har16}.
Other than the stars of M67, there are also two field stars
(HD~187013 and HD~224930) with enhanced activity at large
Rossby numbers of around 2.5, indicative of antisolar
differential rotation; see \cite{BG18}.

\subsection{Stellar surface magnetic field structure}
\label{StellarSurfaceStructures}

Mean-field models have long shown that the surface magnetic field
structure does not always have to be of solar type, i.e., with
a toroidal field that is antisymmetric about the equatorial plane
\citep{Rob71}.
It could instead be symmetric about the equator, i.e., quadrupolar
instead of dipolar.
Yet another possibility is that the large-scale field is nonaxisymmetric,
for example with a dominant azimuthal order of unity \citep{Rae73}.

Early mean-field models of \cite{Rob71} have demonstrated that quadrupolar
mean fields are preferred when the dynamo operates in thin spherical
shells.
In principle, the break point where this happens should be for models that
have convection zones that are somewhat thicker than that of the Sun.
From that point of view, it is unclear why the Sun has an
antisymmetric field and not a symmetric one.
This problem is somewhat reminiscent of the problem of why the Sun has
solar-like differential rotation at the solar rotation rate and not an
antisolar-like, as theoretically expected.
Thus, again, simulations of the solar dynamo seem to place the model
in a position in parameters space that is shifted somewhat relative to
what is theoretically expected.
These two problems may even have a common origin, related, for example,
to the convective conundrum \citep{Lord,CR16,FH16}, i.e., the lack of
power at large length scales in observations relative to the models.
This is possibly explained by stellar convection being dominated by
thin downdrafts or threads which, in the Sun, result from the cooling
near the surface \citep{Spr97}.
This leads to the phenomenon of what is called entropy rain
\citep{Bra16}, where a significant fraction of the energy is being
carried by the Deardorff term; see \Sec{Deardorff}.

Regarding nonaxisymmetry, we do expect rapidly rotating stars to exhibit
nonaxisymmetric magnetic fields.
It is conceivable that the convection can develop spontaneously
a marked nonaxisymmetric modulation, as has been seen in the simulations
of \cite{Bro08}.
This can lead to an $\alpha$ effect that is nonaxisymmetric.
Such models have been studied in the context of galactic dynamos where
such a modulation through the spiral arms is conceivable \citep{MBT91}.
As already discussed in \Sec{HorizontalAveraging}, this implies that the
Reynolds rules cannot be applied.
Not much is known about this case, which deserves further study.

\begin{figure}\begin{center}
\includegraphics[width=.78\columnwidth]{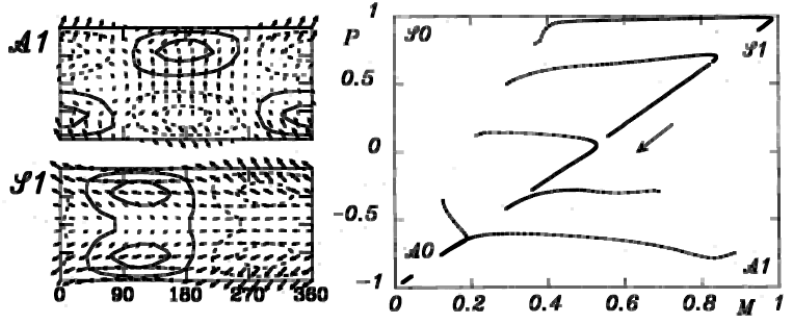}
\end{center}\caption[]{
{\em Left}: Surface magnetic field structure of nonlinear nonaxisymmetric
with an $m=1$ azimuthal order for magnetic fields that are symmetric
(${\cal S}1$) or antisymmetric (${\cal A}1$) about the equatorial plane.
{\em Right}: Evolutionary tracks of solutions in the $MP$ diagram.
Adapted from \cite{RWBMT90}.
}\label{27}\end{figure}

Theoretically, nonaxisymmetric magnetic fields can also be caused by
the $\alpha$ effect becoming anisotropic.
We recall that $\alpha_{ij}$ is a pseudo tensor that can be constructed
from products of terms proportional to gravity $\grav$ (a polar vector)
and angular velocity $\OO$ (an axial or pseudo vector).
The term $\grav\cdot\OO\,\delta_{ij}$ is particularly important because
it leads to $\alpha$ effect dynamo action.
However, there are also terms proportional to $g_i\Omega_j$ and
$g_j\Omega_i$ that were already present in the early work of
\cite{SKR66}.
These are important, because they can favor the generation of
nonaxisymmetric magnetic fields \citep{Rae86a,Rae95}; see the left panel
of \Fig{27} for symmetric and antisymmetric magnetic field configurations
with an azimuthal order of $m=1$.
These solutions are referred to as ${\cal S}1$ and ${\cal A}1$, respectively.
Here, script letters have been used to indicate that these nonlinear
solutions are no longer the same pure composition of modes as in
linear theory.

For rapid rotation, higher powers of $\OO$ are expected, so we
expect a term of the form $\grav\cdot\OOmega\;\Omega_i\Omega_j$,
as was obtained by \cite{Mof72} and \cite{Rue78}.
This term enters with a minus sign and thus tends to cancels the
component $\alpha_{zz}$, where we have assumed that $\OO$ points
in the $z$ direction.
The Roberts flow~I is an example of a flow that has $\alpha_{zz}=0$;
see \Eq{NonlinSat}.
The resulting mean magnetic field has only $x$ and $y$ components,
corresponding to a global magnetic field of that of a dipole
lying in the equatorial plane.

If this should be a model of the geodynamo, 
it is unclear why the Earth's magnetic field is then not also
nonaxisymmetric, given that its Coriolis number is expected to be much
larger than that of many stars.

We have the same problem also for the giant planets Jupiter and Saturn
which have basically axisymmetric magnetic fields, while Uranus and
Neptune are known to have nonaxisymmetric fields corresponding to a
dipole lying in the equatorial plane \citep{RN90}.
A possible explanation for the occurrence of asymmetric mean magnetic
fields in rapid rotators could be the presence of a small but sufficient
amount of differential rotation in Jupiter and Saturn which prevents
the excitation of nonaxisymmetric magnetic fields \citep{Rae86b,Rae95}.
Corresponding mean-field calculations were presented by \cite{MB95}.

Regarding stellar magnetic fields, several stars are seen to have
nonaxisymmetric magnetic fields \citep{Rosen16,See16}.
Those are indeed rapidly rotating stars.
However, the breakpoint between predominantly axisymmetric and
predominantly nonaxisymmetric magnetic fields is observed to be at about
5 times the solar rotation rate \citep{Lehtinen16}, while simulations
suggest this to happen already at about 1.8 times the solar value
\citep{Viviani18}.

When the anisotropy is weak, the axisymmetric dipole solution ${\cal A}0$
is often the most preferred one.
Nevertheless, even in that case the nonaxisymmetric ${\cal S}1$ solution
can occur as a transient for an extended period of time, if the initial
condition has a strong symmetric component.
As shown in a state diagram (\Fig{27})
of parity $P$ ($=1$ for symmetric and $-1$
for antisymmetric fields) versus nonaxisymmetry $M$ (i.e., the fractional
energy in the nonaxisymmetric components), the solution first evolves
to become more symmetric with respect to the equatorial plane ($P\to1$),
but more nonaxisymmetric ($M\to1$), until it evolves along the diagonal
in the $PM$ diagram toward the ${\cal A}0$ solution \citep{RWBMT90}; see
the right panel of \Fig{27}.
If only axisymmetric solutions are permitted, the ${\cal S}0$ solution
would be a stable end state \citep{BKMMT89}.
However, as was shown by \cite{RW89}, this is an artifact of the
restriction to axisymmetry.
Fully nonaxisymmetric models demonstrate that the stellar surface field
can undergo extended transients via a nonaxisymmetric mode before the
axisymmetric dipole solution is restored.
This could potentially be important in understanding the nature of
the secondary cycles observed in stellar dynamos; see \cite{BMM17}.

\section{Accretion disk dynamos}

Unlike stars, accretion disks are flat.
Early simulations in the context of galactic dynamos have suggested for
some time that the toroidal magnetic fields in disks should be symmetric
about the midplane, i.e., quadrupolar \citep{RSS88,BBMSS96}.
This was indeed confirmed by the first simulations of magnetic
fields generated by turbulence from the magneto-rotational instability
\citep{BNST95,HGB96,SHGB96}.

\subsection{Unconventional sign of $\alpha$}

The early simulations of \cite{BNST95} indicated that accretion
disks have an $\alpha$ effect that is negative in the upper disk plane,
which was rather unexpected.
Here, $\alpha$ was measured simply by correlating the local toroidal
value of $\meanEMF$ (corresponding to $\meanemf_y$ in their shearing
box simulations) with the mean toroidal magnetic field (corresponding
to $\meanB_y$).
Similar results were later reproduced by \cite{ZR00}.
As explained \Sec{TFM}, this method is not always reliable.
Nevertheless, subsequent simulations with the test-field method
have confirmed that the relevant component $\alpha_{yy}$ is
negative \citep{B05QPO,GZER08}, at least close to the midplane
\citep{GZER08,Gre13,GP15}.

Local mean-field models with a negative $\alpha$ effect in the upper disk
plane predicted oscillatory magnetic fields \citep{Bra98}, which agrees
with what is seen in the simulations of \cite{BNST95}.
Again, however, \cite{Gre13} and \cite{GP15} found that the sign
may change in the outer parts, where they found it to be the usual one,
i.e., positive in the upper disk plane.

The theoretical explanation for an unconventional sign could be
related to a dominance of a magnetic buoyancy-driven $\alpha$ effect;
see \cite{BS98} for numerical results in the context of stellar dynamos.
The idea is that a magnetic field that is enhanced locally in a flux tube
leads not only to its rise, but also to its contraction along the tube
\citep{BC97}.
If this effect dominates over the expansion of rising gas, it could explain
the opposite sign of $\alpha$.
This could indeed be the right explanation \citep{RP00,ZR00}.
Magnetically driven turbulence might also be relevant to the
Sun and could cause unconventional turbulent transport \citep{RPB01,CMRB11}.

\subsection{Identifying $\alpha\Omega$-type dynamo action in disk simulations}
\label{IdentifyingAlphaOmega}

\begin{figure}\begin{center}
\includegraphics[width=.9\columnwidth]{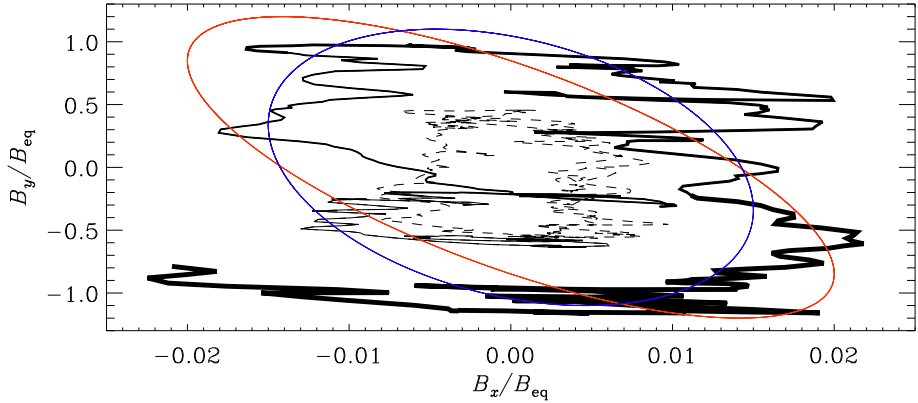}
\end{center}\caption[]{
Phase plot of the averages of poloidal and toroidal fields
over narrow slices in the $z$ direction of the simulation domain.
Early times are plotted as dashed lines, but later
times are solid with increasing thickness toward the end
showing that a point on the curve moves forward in a clockwise
direction.
Overplotted are two ellipses showing $B_{y0}\propto\cos(\omega t+\phi)$ versus
$B_{x0}\propto\cos\omega t$ with (i) $\phi=0.6\pi$, $B_{x0}=0.015\,B_{\rm eq}$,
and $B_{y0}=1.1\,B_{\rm eq}$ (blue line) and (ii) $\phi=0.75\pi$,
$B_{x0}=0.02\,B_{\rm eq}$, and $B_{y0}=1.2\,B_{\rm eq}$ (red line).
Note that (i) fits slightly better than (ii), but both fit poorly
in the lower left quadrant.
}\label{phase2_64x256c8_test}\end{figure}

To identify $\alpha\Omega$-type dynamo action
as the main course of oscillations seen in simulations,
it is advantageous to determine the phase relation between
poloidal and toroidal fields \citep{Bra08}.
This is a standard tool in solar dynamo theory for inferring the
sense of radial differential rotation.
Mean-field theory predicts a phase shift by $3\pi/4$.
Simulation results, however, are suggest a somewhat smaller phase
shift of $0.6\pi$; \Fig{phase2_64x256c8_test}.

An alternative idea is magnetic buoyancy being the reason for migration
away from the midplane \citep{SSAB16}.
However, no detailed proposal for the phase relation from the buoyancy
effect has yet been made.
By comparison, the interpretation of the magnetic field migration in terms
of an $\alpha\Omega$ dynamo is rather straightforward; see \cite{GP15}
for a recent analysis.

\subsection{Incoherent $\alpha$--shear dynamo}
\label{IncoherentAlphaShear}

It has been suggested that the magnetic field of accretion disks
could be explained by what is known as an incoherent
$\alpha$--shear dynamo \citep{VB97}.
This type of effect is a hybrid between a fluctuation dynamo
(i.e., small-scale dynamo) and a mean-field dynamo and involves
fluctuations in the mean field itself.
The occurrence of fluctuations in the mean field is a natural outcome
of finite scale separation when the turbulent eddies are comparable to
the size of the domain along the direction of averaging.
This was originally proposed by \cite{Hoyng,Hoyng93} to explain
irregularity of standard $\alpha\Omega$ dynamos.
He discussed the occurrence of fluctuating mean fields, but not the
occurrence of a new mean-field dynamo effect.
The occurrence of a new dynamo effect is possible when there is also
strong differential rotation together with turbulent diffusion \citep{VB97}.
The verification of this mechanism from simulations was discussed in
\cite{BRRK08}, who measured $\alpha(z,t)$ and found that its rms value,
$\bra{\alpha^2}^{1/2}$, was large enough to explain the dynamo action
found in their model.
Unlike $\alpha\Omega$ dynamos, which rely on the presence of stratification
to produce an $\alpha$ effect, this is not required for the incoherent
$\alpha$--shear dynamo effect.
\cite{You08a,You08b} have suggested instead a mechanism which they called
a shear dynamo.
It is not clear, however, whether this is really a new mechanism, but
several similarities with the incoherent $\alpha$--shear dynamo effect
such as the linear scaling of the growth with the shear rate have been
pointed out \citep{Pro07,HMWS11,MB12}.

\subsection{The shear--current effect}

There is also the possibility of a dynamo effect from what is known as
the shear--current effect \citep{RK03,RK04}.
There is, however, no independent verification of this effect
\citep{B05QPO,RK06,RS06}.
\cite{SS09} found this term to vanish under SOCA.
Subsequent work by \cite{SB15,SB16} has shown that this effect may work
when there are small-scale magnetic fields, for example those
produced by small-scale dynamo action.
In their first paper, \cite{SB15} demonstrated this effect using
magnetic forcing, which is known to lead to potentially peculiar results
that are not in any known relation to those in naturally occurring
hydromagnetic turbulence; see \cite{RB10} for their calculations of
turbulent transport coefficients in kinetically and magnetically driven
flows.
However, in their later work \citep{SB16}, magnetic fluctuations resulted
entirely from the small-scale dynamo effect.
To gain more faith in the reliability of their results, it would
be of interest to verify their results using the fully nonlinear
test-field method.
\cite{SSH16} have shown that the shear--current effect
could, with suitably adjusted parameters, reproduce the magnetic
cycles rather well.

However, to find conclusive evidence for a magnetic version of
this effect, as advocated by \cite{SB16}, one needs to apply the fully
nonlinear version of the test-field method to such simulations.
It would be important to verify that this fully nonlinear method
is indeed required in cases where the small-scale dynamo is excited.
So far, however, no such evidence has been presented yet; see \cite{RB10}
for a corresponding discussion.

\subsection{Magnetic Prandtl number dependence}

At about the same time when it became clear that small-scale
dynamos are harder to excite at small values of the magnetic Prandtl
number \citep{Scheko05}, it was noticed that dynamos driven by the
magneto-rotational instability are no longer excited at small magnetic
Prandtl numbers \citep{FP07,FPLH07}.
This may indeed be for the same reason that the small-scale dynamos
become harder to excite.
It still needs to be demonstrated, then, that at larger magnetic
Reynolds numbers, the dynamos become excited again.

The assumption of periodic boundary conditions in simulations of
the magneto-rotational instability is crucial for obtaining the result
that those dynamos are no longer or not that easily excited at small
magnetic Prandtl numbers.
Comparisons by \cite{KK11} with the vertical field (pseudo vacuum)
condition on the upper and lower boundaries have shown that the dynamo
is no longer dependent on the microphysical value of the magnetic
Prandtl number.
This was interpreted as a consequence of large-scale dynamo action
being possible in this case.
This dynamo might well be the incoherent $\alpha$--shear dynamo;
see \Sec{IncoherentAlphaShear}.

\section{Galactic dynamos}

The realization that interstellar space harbors magnetic fields has
intrigued scientists already in the 1950s \citep{BS51} and the idea of
a turbulent origin was anticipated \citep{Bat50}.
His early theory of what is nowadays called a small-scale dynamo was a
simple one, but it turned out to be incorrect and was later superseded
by the work of \cite{Kaz68}; see also \cite{RK97} for the generalization
of this theory to finite magnetic Prandtl numbers.
The application of mean-field theory started with the work of \cite{VR71}
and \cite{Par71a}.

\subsection{The $\alpha$ effect in galactic turbulence}

Galactic dynamos are similar to accretion disk dynamos in that their
geometry is flat, but here, turbulence and thus an $\alpha$ effect can
be driven by supernova explosions \citep{Fer92a,Fer92b,Fer93a,Fer93b}.
Those calculations showed an unexpected result in that the vertical
component of the $\alpha$ tensor was negative in the northern hemisphere;
see \cite{Fer93a}.
This unusual sign of $\alpha_{zz}$ was first found in convection
simulations \citep{BNPST90}.

Of course, $\alpha_{zz}$ can only be determined if one allows for
vertical mean magnetic fields.
This was done in \cite{BRK12}, where a special test-field method for
axisymmetric turbulence was adopted.
However, under the physical conditions considered (stably stratified
rotating turbulence), the sign of $\alpha_{zz}$ was found to be
mostly the same as for the horizontal $\alpha$ effect; see their
figure~8, where only for $\Rm\approx40$ a negative value was found
($\alpha_{zz}=0.002\,\urms/3$, which is rather small).

\subsection{Capturing the galactic dynamo effects in numerical simulations}

Simulations by \cite{BKMLM99} where the first ones that produced
small-scale dynamo action in the interstellar medium.
The first ones showing large-scale dynamo action were those by
\cite{GEZR08}, but that was at four times the actual rotation rate.
Interestingly, \cite{KGVS18} found a near cancellation of the total net
helicity from the contributions produced by rotation and shear with
opposite signs.
This may explain the difficulties encountered by \cite{GEZR08} in getting
the large-scale dynamo excited at the actual rotation rate.

The simulations of \cite{GZER08} produced detailed predictions for
the tensors $\alpha_{ij}$ and $\eta_{ijk}$ using the test-field method.
Contrary to the results of \cite{Fer92b}, they found that turbulent
pumping is directed toward the midplane, as was already assumed in
\cite{BDMSS93}.
The simulations of \cite{Gent13a} were the first to produce
large-scale dynamos for the actual values of the galactic rotation rate.
They also found small-scale dynamo action, but their Prandtl
number was varying between the different structural phases generated.
This is because the viscosity was set proportional to the sound speed,
hence it was very large in the hot phase and very small in the cold phase.
A constant magnetic diffusivity was used on top of this, resulting in
large $\Pm$ in the hot phase, and hence more favorable conditions for
small-scale dynamo action.
Therefore, the interpretation of those results is not obvious.

\subsection{Axisymmetric and bisymmetric spirals: significance of the arms}

An obvious question concerns the importance of spiral arms
in making the $\alpha$ effect nonaxisymmetric and thus causing or
facilitating nonaxisymmetric magnetic fields.
The perhaps only galaxy where nonaxisymmetric magnetic fields have
been detected is M81, while the magnetic field detected in many other
galaxies are predominantly axisymmetric; see \cite{BBMSS96}.
\cite{MS91} found that the $m=1$ mode could grow if $\alpha$
is assumed to be nonaxisymmetric.
\cite{CSS13} extended these considerations to try and explain magnetic
spirals, also using the time nonlocality of mean-field dynamo theory
(see \Sec{NonlocalityScaleSeparation}).
Simulations with a nonaxisymmetric $\alpha$ effect have shown that the
marginal dynamo numbers for nonaxisymmetric dynamos are substantially
lowered when the $\alpha$ effect is nonaxisymmetric \citep{MBT91}.
It is not obvious, however, that the magnetic field coincides with the
gaseous arms and there are arguments that magnetic and gaseous arms are
actually interlaced \citep{Shu98}.

\subsection{Significance of galactic halos}

Galaxies also have extended halos that could support dynamo action.
The main difference between dynamos in the disk and in the halo is
that halo dynamos behave essentially like stellar ones in that they
are expected to produce a dipolar magnetic field whereas the disk dynamo
is expected to produce a quadrupolar magnetic field.
This can lead to interesting interactions between the two
\citep{BKMMT89,SS89}.
The occurrence of mixed modes between symmetric and antisymmetric
fields was first proposed by \cite{SS90} and then tested numerically
by \cite{BDMSS92}.
It has also been proposed that the galactic bulge may provide another
near-spherical entity that could harbor dipolar magnetic fields
\citep{DB90}.

An important question concerns the direction of turbulent pumping.
Is it directed toward the disk midplane or away from it?
\cite{BDMSS93} discussed the possibility that it is directed
toward the disk midplane, which could lead to an
enhancement of the dynamo effect by making the field more concentrated.
This was indeed supported by the simulations of \cite{GZER08,GEZ13}.

At large radii, sufficiently far away from the galactic center
where supernova explosions no longer occur and supernova driving
becomes inefficient, the magneto-rotational instability
could also act in the galaxy \citep{SB99}.
This idea has been explored in a number of subsequent papers.
The work of \cite{PO07} showed that the thermal instability interacts
with the turbulence from the magneto-rotational instability to produce
a network of cold filamentary clouds embedded in a warm diffuse ambient
medium.
\cite{KKV10} found that the stresses from the magneto-rotational
instability become strongly suppressed with increasing forcing.
In the simulations of \cite{MNKASM13}, magnetic flux escapes from the
disk by the Parker instability and drives dynamo activity
by generating disk magnetic fields with opposite polarity.
The subsequent amplification of a disk magnetic field by the
magneto-rotational instability causes quasi-periodic reversals of
azimuthal magnetic fields on a timescale of ten rotation periods.
\cite{BGE15} also found that vertical-flux initial conditions are able to
influence the galactic dynamo via the occurrence of the magneto-rotational
instability.

\subsection{Cosmic ray driven turbulence}
\label{CosmicRay}

In modelling the galactic dynamo, an additional energy source is provided
by cosmic rays, which can inflate magnetic flux tubes and thus make
them buoyant, which causes them to rise and thereby exert work on the
magnetic field.
This was first addressed by \cite{Par92} and has been modelled numerically
by \cite{HKOL04,HOKL09} in local models and by \cite{HWK09} in global
models.
It has even been argued that the presence of cosmic rays helps to make
the galactic dynamo ``fast,$\!\!$'' i.e., independent of the microphysical
resistivity.
This question remains somewhat puzzling, because one would have thought
that {\em any} turbulent dynamo would be a fast one, at least in the
kinematic sense, because the kinematic values of $\alpha$ and $\etat$
are thought to be independent of the microphysical value of $\eta$.
This is also confirmed by numerical simulations \citep{SBS08,BRRS08,BSR17}.
Given that the cosmic ray diffusivity is very large, \cite{SBMS06}
used in their simulations a non-Fickian telegrapher's equation approach
discussed in \Sec{TheTrick}.

In the scenario discussed above, cosmic rays inflate magnetic
field structures and make them buoyant in an external gravity field.
This is not the most direct way of cosmic rays driving turbulent motions.
Another process is to invoke the electric current associated with the
flow of protons in the cosmic rays.
If there is a magnetic field with a component aligned with this
current, it can drive an instability \citep{Bell04}.
This can lead to turbulence and a slow continued build-up of magnetic
field in terms of $\alpha$ effect.  
Furthermore, since the magnetic field and the current density form a
pseudo-scalar, it is not surprising that their presence causes a turbulent
$\alpha$ effect that explains the slow growth of the magnetic field
after the initial exponential phase is over.
\cite{BL14} measured the anisotropy of such Bell turbulence
and found $\ell^{2/3}$ and linear scalings of the perpendicular
and parallel second order structure functions, as also expected for
regular hydromagnetic turbulence \citep{GS95}.

\subsection{Mode cleaning by nonlinearity}

Even though the kinematic dynamo may be a fast one, as discussed in
\Sec{CosmicRay}, it may not be sufficiently prominent owing
to the dominance of small-scale dynamo action \citep{Beck94}.
There is work suggesting that large-scale dynamos work successfully only
because of nonlinearity \citep{CH09}.
This notion was already supported by the work of \cite{Bra01}, which showed
that in the kinematic regime, no large-scale field was found and that
it was only near the end of the nonlinear phase that large-scale magnetic
fields became fully developed.
This can also be seen by looking at \Fig{B}.

One reason for the emergence of a large-scale field only in the nonlinear
phase is the fact that there can be multiple solutions to the large-scale
dynamo problem: not only can a large-scale field develop in any of the
three coordinate directions, but, in a periodic domain, it can also come
with any possible phase shift.
Also, if the scale separation is large, the direction of the large-scale
does not need to be any of the coordinate directions, and many of the
intermediate directions are possible.
This explains the extended time interval during which large-scale,
but incoherently arranged patches of magnetic field are present;
see \Fig{B} of \Sec{CatastrophicQuenching}.
\cite{SB14} have shown that the kinematic dynamo does operate in
high Reynolds number turbulence and that one really has a new type
of dynamo that has aspects of small-scale and large-scale dynamos.
Interestingly, as the dynamo saturates, even the small-scale fields
attain more power at intermediate length scales \citep{PB12,BSB16}.

\section{Early Universe}

The connection between the early Universe and mean-field dynamos is not
evident, because no mean fields have ever been observed and such fields
are also not really expected.
Instead, we expect a turbulent magnetic field.
On the other hand, the possibility that a turbulent magnetic field might
have helicity has frequently been discussed \citep{BEO96,CHB01,FC02}.
The most important reason is that then a turbulent magnetic field can
undergo efficient inverse cascading \citep{PFL76}, which significantly
increases the turbulent correlation length of the magnetic field from
the scale of a few centimeters at the time of the electroweak phase
transition to about $10^8\cm$, which, after the cosmological expansion
of the Universe, would correspond to about $30\kpc$, making it a strong
candidate for explaining the large-scale magnetic fields in the Universe
\citep{BJ04,KTBN13}.

\subsection{Inversely cascading turbulent magnetic fields}

There are lower limits on the strength of a diffuse magnetic field
throughout all of space of about $10^{-14}$ to $10^{-18}\G$ on a scale
of about $1\Mpc$ \citep{Aharonian,TVN11,Dermer}.
These limits constrain the product of magnetic energy and length
scale, $\bra{\BB^2}\xiM$, so the lower limit would be ten times larger
if $\xiM$ was a hundred times smaller.
On dimensional grounds, this product can also be a measure of the modulus
of the magnetic helicity \citep{BSRKBFRK17}.

Simulations have shown that the magnetic energy spectra $\EM(k,t)$ of
decaying turbulence tend to display a selfsimilar behavior \citep{BK17},
\EQ
\EM(k,t)=\xiM^{-\beta}\phiM\left(k\xiM(t)\right).
\EN
where $\xiM$ is the magnetic correlation length, $\phi_{\rm M}$ is a
universal function for the magnetic spectra at all times, and $\beta$
is an exponent that depends mostly on the physics governing the decay
and, in some cases, also on the initial conditions \citep{Ole97}.
For example, $\beta=0$ in the fully helical case when $\bra{\AAA\cdot\BB}$
is conserved, $\beta=1$ when $\bra{\AAA^2}$ is conserved, $\beta=2$ when
the Saffman integral is conserved, and $\beta=4$ when the Loitsiansky
integral is conserved; see \cite{BK17} for details.

Assuming that $\xiM(t)\propto t^q$ with exponent $q$, we then expect the
magnetic energy to decay like
\EQ
\EEM(t)=\int_0^\infty\EM(k,t)\,\dd k
=\xiM^{-(\beta+1)}\int_0^\infty\phiM(k\xiM)\,\dd(k\xiM)
\propto t^{-(\beta+1)q}\propto t^{-p},
\EN
so $p=(\beta+1)q$ is the exponent on the decay of magnetic energy.
Furthermore, as noted by \cite{Ole97}, the hydrodynamic and hydromagnetic
equations are invariant under rescaling, $x\to\tilde{x}\ell$ and
$t\to\tilde{t}\ell^{1/q}$, which implies corresponding
rescalings for velocity $u\to\tilde{t}\ell^{1-1/q}$ and viscosity
$\nu\to\tilde{\nu}\ell^{2-1/q}$.
Furthermore, using the fact that the dimensions of $E(k,t)$ are given by
$[E]=[x]^3[t]^{-2}$, and requiring $\phiM$ to be invariant under rescaling
$E\to\tilde{E}\ell^{3-2/q}\propto\tilde{k}^\beta\ell^{-\beta}\psi$,
he finds that $\beta=-3+2/q$.
This is indeed compatible with simulations of nonhelical hydromagnetic
turbulence \citep{Zra14,BKT15}.

\subsection{Connection with mean-field theory}

The helical decay law has been modelled using mean-field theory for
the spectra $\EM(k,t)$ and $\HM(k,t)$ in the form \citep{Cam07}
\EQ
{\partial\EM\over\partial t}=-2(\eta+\etat)k^2\EM+\alpha k^2\HM,
\label{Cam07a}
\EN
\EQ
{\partial\HM\over\partial t}=-2(\eta+\etat)k^2\HM+4\alpha \EM,
\label{Cam07b}
\EN
where $\etat$ and $\alpha$ are here time-dependent coefficients
with $\etat=\tau_{\rm d}\int\EM\,\dd k$ being the magnetic diffusivity and
$\alpha=\tau_{\rm d}\int k^2\HM\,\dd k$ is a purely magnetic contribution to the
$\alpha$ effect.
The assumption of \cite{Cam07} that $\etat$ can, in this case of strong
magnetic fields, be assumed to be proportional to the magnetic energy
density needs to be verified, as it would seem to contradict the results
form the second order correlation approximation in the kinematic case,
as discussed at the end of \Sec{DoesMagnetic}.
The timescale $\tau_{\rm d}$ is assumed constant in these considerations
and equal to the friction or drag time that is introduced when replacing
the nonlinear term $\uu\cdot\nab\uu$ by $\uu/\tau_{\rm d}$.
This approximation was already used by \cite{Sub99} who referred to it
as the ambipolar diffusion nonlinearity.
\cite{BS00} solved his model numerically and also obtained inverse
cascading.

\begin{figure}\begin{center}
\includegraphics[width=\columnwidth]{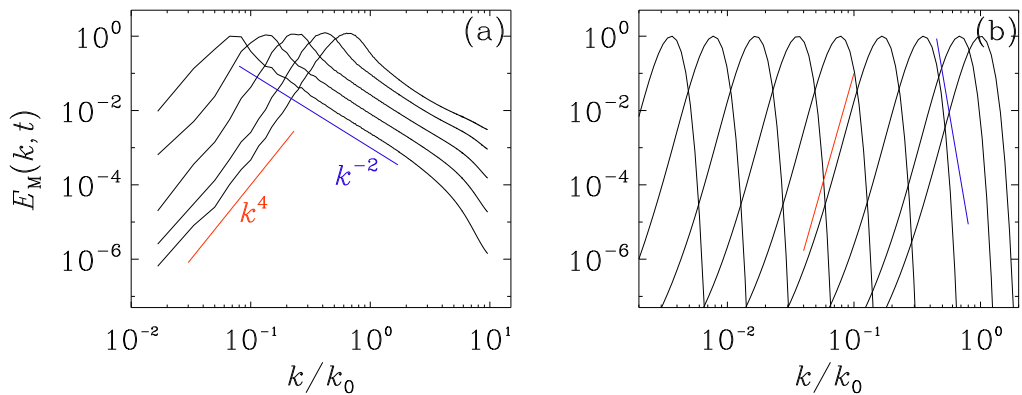}
\end{center}\caption[]{
(a) Fully helical three-dimensional turbulence simulation of
a decaying initially fully helical turbulent magnetic field.
The velocity is driven entirely by the Lorentz force of the magnetic field.
The time in units of the initial Alfv\'en time are 17, 50, 150, 430, and 1200.
The red and blue lines are proportional to $k^4$ and $k^{-2}$, respectively.
(b) Solution of \eqs{Cam07a}{Cam07b} shown at times 1, $10^2$,
$10^3$, ..., until $10^9$.
The red and blue lines are proportional to $k^{12}$ and $k^{-20}$,
respectively.
}\label{pmodel}\end{figure}

The solutions to these equations characterize certain aspects of the
helical decay law, but they do not correctly describe details of the
spectra, as shown in \Fig{pmodel}.
In particular, the model does not reproduce the $k^4$ subinertial
range spectrum \citep{DC03} and also not the $k^{-2}$ inertial range
\citep{BKT15}.

\subsection{Comments on the chiral magnetic effect}

The equations have been generalized to the case where magnetic
helicity can be generated through what is known as the chiral magnetic
effect.
This is an effect of relativistic fermions whose spin aligns
with the magnetic field, leading to oppositely oriented currents from
left- and right-handed fermions.
At low temperatures, the spin can flip rapidly, so there is no net current,
but this is not the case under relativistic conditions.
In that case, when the difference in the number densities between left-
and right-handed fermions, i.e., their chemical potential,
is different from zero, it leads to a field-aligned current
proportional to $\mu\BB$.
This is {\em formally} equivalent to an $\alpha$ effect, although it is
here not connected with turbulence, but it is a microphysical effect
\citep{JS97,BFR12,BFR15,Roga17,Schober18}.
The total chirality is however conserved, so
$\mu+\half\lambda\bra{\AAA\cdot\BB}=\const\equiv\mu_0$,
i.e., it is equal to the initial chemical potential $\mu_0$ if the
initial magnetic helicity was vanishing.
This implies that a fully helical magnetic field can be produced by
exponential amplification from a weak seed magnetic field.
This continues until the magnetic helicity (multiplied by $\lambda/2$)
reaches the value $\mu_0$ at later times.
Similar to the simulations without the chiral magnetic effect, the
difference between the two models is related to the absence of a forward
cascade \citep{DS17,PLS17,BSRKBFRK17}.

\section{Conclusions}

The applications of mean-field theory to astrophysical bodies has been
far from straightforward.
One might have thought that, given that so much is known about the
expressions for $\alpha_{ij}$ and $\eta_{ijk}$, and that even the
inclusion of nonlocality is now straightforward, it should not be a problem
to apply the full theory to the Sun or to galaxies.
This is true in theory, and some models of galactic and solar
dynamos now include nonlocality in space and/or time; see \cite{CSS13}
and \cite{BC18}, respectively.
In practice, however, success remained limited because it looked like that
models for the Sun did not reproduce the Sun too well.
It was therefore though that this problem could be fixed by
``massaging'' some of the coefficients such that the model works, but even
that did not seem to lead to satisfactory results.
In the wake of this type of experience, the flux transport model was
developed, which was not just a refinement of theoretically justified
models, but it was guided entirely by the desire to make the model work
for the Sun.
This remains unsatisfactory even today.
The problem with this is that, given that such a flux transport dynamo
has no theoretical basis, it is unclear whether such a model can be
applied in a predictive manner to other stars.
In that respect, it was already noted that the flux transport dynamo
does not seem to be able to explain the rising branches seen in
\Fig{pCycleOutput}, but only a declining branch obtained by fitting one
line through both branches \citep{JBB10,KKC14}.

Alternatively, one may argue that the solar dynamo simply cannot be
treated with mean-field theory, and that we just have to wait for numerical
simulations to resolve the Sun sufficiently well in space and time to
reproduce its main features such as the equatorial migration or the
toroidal flux belts, spoke-like angular velocity contours, and
the near-surface shear layer.
While this viewpoint may turn out to be true in the end, the argument for
this remains unsatisfactory simply because we clearly do see a well-defined
mean field with large-scale spatial and temporal order.
Therefore, there is {\em a priori} no reason why there should be
no theory for describing such a mean field, which clearly does seem to exist.
On the other hand, it is true that the full range of mean-field coefficients
and effects can be rather large and too complex to be dealt with in
a fully predictive manner without fudge parameters.
Thus, mean-field theory might in principle still be correct, but
impractical under conditions of practical interest.

This unsettled situation is obviously one of the reasons why---after all
these years---mean-field theory is still a very active field of research,
and thus it is the very reason for having this special issue in JPP.

\begin{acknowledgements}
I feel privileged to acknowledge Karl-Heinz R\"adler for having taught
me so many important aspects of dynamo theory through our joint work
since we first met in Helsinki in 1986.
I also thank the two referees for their detailed and constructive reports.
This research was supported in part by the NSF Astronomy and Astrophysics
Grants Program (grant 1615100), and the University of Colorado through
its support of the George Ellery Hale visiting faculty appointment.
\end{acknowledgements}


\end{document}